\documentclass[aps,showpacs,twocolumn]{revtex4}

\usepackage{graphicx}

\newcommand{\bphi}{{\mbox{\boldmath $\varphi$}}}
\newcommand{\bomega}{{\mbox{\boldmath $\omega$}}}

\begin{document}

\title{Multi-level quantum description of decoherence in superconducting qubits}

\author{Guido Burkard}
\author{Roger H. Koch}
\author{David P. DiVincenzo}
\affiliation{IBM T.\ J.\ Watson Research Center,
         P.\ O.\ Box 218,
         Yorktown Heights, NY 10598}

%\date{\today}

\begin{abstract}
We present a multi-level quantum theory of decoherence for a general
circuit realization of a superconducting qubit.
Using electrical network graph theory, we derive a Hamiltonian 
for the circuit.  The dissipative circuit elements (external impedances,
shunt resistors) are described using the Caldeira-Leggett model.
The master equation for the superconducting phases
in the Born-Markov approximation is
derived and brought into the Bloch-Redfield form in order to 
describe multi-level dissipative
quantum dynamics of the circuit.  The model
takes into account leakage effects, i.e.\ transitions from the allowed qubit 
states to higher excited states of the system.
As a special case, we truncate the Hilbert space and derive a 
two-level (Bloch) theory with characteristic relaxation ($T_1$) 
and decoherence ($T_2$) times.  
We apply our theory to the class of superconducting flux qubits;
however, the formalism can be applied for both superconducting 
flux and charge qubits.
\end{abstract}

\pacs{03.67.Lx, 74.50.+r, 85.25.Dq, 85.25.Cp, 72.70.+m}
% PACS numbers:
% 03.67.Lx - Quantum Computation
% 74.50.+r Tunneling phenomena; point contacts, weak links, Josephson effects
% 85.25.Dq Superconducting quantum interference devices (SQUIDs)
% 85.25.Cp Josephson devices
% 72.70.+m Noise processes and phenomena

\maketitle

\section{Introduction}
\label{introduction}

Since the famous cat paradox was formulated by Schr\"odinger \cite{Schroedinger}, 
the question whether the range of validity of quantum mechanics in principle extends
to macroscopic objects has been a long-standing open problem.
While macroscopic quantum tunneling was observed in several experiments
\cite{VossWebb,Martinis87,Clarke88,Rouse95}, there is less experimental
evidence for macroscopic quantum coherence.
The experimental study of macroscopic superconducting circuits comprising
low-capacitance Josephson junctions as a physical implementation of a 
quantum computer (see Ref.~\onlinecite{MSS} for a review)
represents a new test for macroscopic quantum coherence.
On the theory side, the effect of dissipation on macroscopic quantum tunneling and
macroscopic quantum coherence was put into a quantitative phenomenological model by
Caldeira and Leggett \cite{CaldeiraLeggett}.

The fundamental building block of a quantum computer \cite{NC}
is the quantum bit (qubit)--a quantum mechanical two-state system
that can be initialized, controlled, coupled to other qubits, and
read out at the end of a quantum computation.
Presently, three prototypes of superconducting qubits are studied experimentally.  
The charge ($E_C\gg E_J$) and the flux ($E_J\gg E_C$) 
qubits are distinguished by their Josephson junctions' relative 
magnitude of charging energy $E_C$ and Josephson energy $E_J$.
A third type, the phase qubit \cite{Martinis02}, 
operates in the same regime as the flux qubit, but it consists of 
a single Josephson junction.  In all of these systems, the quantum state
of the superconducting phase differences across the Josephson
junctions in the circuit contain the quantum information, i.e., the
state of the qubit.
Since the superconducting phase is a continuous variable as, e.g.,
the position of a particle, superconducting qubits (two-level systems)
have to be obtained by truncation of an infinite-dimensional Hilbert 
space.  This truncation is only approximate for various reasons;
(i) because it may not be possible to prepare the initial state 
with perfect fidelity in the lowest two states, (ii) because
of erroneous transitions to higher levels (leakage effects) 
due to imperfect gate operations on the system, and (iii) because 
of erroneous transitions to higher levels due to the unavoidable
interaction of the system with the environment.
One result of the present work is a quantitative estimate of the effect of
errors of type (iii) by studying the \textit{multi-level}
dynamics of a superconducting circuit containing dissipative elements.
The multilevel dynamics and leakage in superconducting qubits may be related
to the observed limited visibility of coherent oscillations.
Previous theoretical works on the decoherence of superconducting
qubits \cite{Tian99,Tian02,WWHM,Wilhelm} have typically relied on the
widely used spin-boson model that \textit{postulates} a purely two-level
dynamics, therefore neglecting leakage effects.
Ref.~\onlinecite{Tian02} includes the dynamics of an attached
measurement device, thus going beyond the standard spin-boson
model while still making the \textit{a priori} two-level assumption.

In this paper, we present a general multi-level 
quantum theory of decoherence in
macroscopic superconducting circuits and apply it
to circuits designed to represent flux qubits,
i.e.\ in the regime $E_J\gg E_C$.  However, the same
formalism can be applied to charge qubits.
Flux qubits have been proposed and studied experimentally
by several groups \cite{Mooij,Orlando,vanderWal,Chiorescu,Friedman,IBM}.
The first step in our analysis is the derivation of a
Lagrangian and Hamiltonian from the classical dynamics of a
superconducting circuit;  the Hamiltonian is then used 
as the basis of our quantum theory of the superconducting
circuit.
While deriving the Lagrangian and Hamiltonian of a dissipation-free
electrical circuit is--at least in principle--rather straightforward, 
different possible representations of dissipative elements 
(such as resistors) can be found in the literature.
One possibility is the representation of resistors as 
transmission lines \cite{YurkeDenker,Yurke,WernerDrummond},
i.e.\ an infinite set of dissipation-free elements
(capacitors and inductors).   Here, we use a related but different
approach following Caldeira and Leggett
by modeling each resistive element by a bath
of harmonic oscillators that are coupled to the
degrees of freedom of the circuit \cite{CaldeiraLeggett,Esteve,Devoret}
(see also Refs.~\onlinecite{LeggettRMP,Weiss} for extensive reviews).

We develop a general method for deriving
a Hamiltonian for an electrical circuit containing 
Josephson junctions using network graph theory \cite{Peikari}.
A similar approach, combining network graph theory with the
Caldeira-Leggett model for dissipative elements, was
proposed by Devoret \cite{Devoret}.
On a more microscopic level, circuit theory was also 
used in combination with Keldysh Green functions
in order to obtain the full counting statistics 
of electron transport in mesoscopic systems \cite{Nazarov}.
Here, we give explicit general expressions for the
Hamiltonian in terms of the network graph parameters
of the circuit.  We apply our theory to Josephson
junction networks that are currently under study as
possible candidates for superconducting realizations
of quantum bits.  
By tracing out the degrees of freedom of the dissipative
elements (e.g., resistors), we derive a generalized master equation
for the superconducting phases.  In the Born-Markov
approximation, the master equation is cast into the
particularly useful form of the Bloch-Redfield equations \cite{Redfield}.
Since we do not start from a spin-boson model, 
we can describe multi-level dynamics and 
thus leakage, i.e.\ transitions from the allowed qubit 
states to higher excited states of the superconducting system.
As a special case, we truncate the Hilbert space and derive a 
two-level (Bloch) theory with characteristic relaxation ($T_1$) 
and decoherence ($T_2$) times.

\section{Overview and Results}
\label{overview}

Before presenting a formal derivation, we explain
the main results and show how they can be applied to calculate the
relaxation, decoherence, and leakage times $T_1$, $T_2$, and $T_L$ 
of a superconducting qubit.
Our theory is capable of predicting more than these quantities
since it can be used to model the evolution of the entire density matrix.
However, we concentrate on the relaxation, decoherence, and leakage time
in order to keep the discussion simple.  For concreteness, we 
discuss the IBM qubit \cite{IBM}, which is described by the 
electrical circuit drawn in Fig.~\ref{ibm-graph}.
The procedure is as follows.
\begin{enumerate}
\item 
Draw and label a \textit{network graph} of the superconducting circuit,
in which each two-terminal element (Josephson junction,
capacitor, inductor, external impedance, current source)
is represented as a branch connecting two nodes.
In Fig.~\ref{ibm-graph}, the IBM qubit is represented
as a network graph, where thick lines are used as a shorthand 
for RC-shunted Josephson junctions (see Fig.~\ref{rsj}).
A convention for the direction of all branches has to be chosen--in
Figs.~\ref{ibm-graph} and \ref{rsj}, the direction of branches
is represented by an arrow.

\item
Find a \textit{tree} of the network graph.   A tree of a graph is
a set of branches connecting all nodes that does not contain
any loops.  Here, we choose the tree such that it contains
all capacitors, as few inductors as possible, and neither resistors 
(external impedances) nor current sources (see Sec.~\ref{ecircuits}
for the conditions under which this choice can be made).
The tree of Fig.~\ref{ibm-graph} that will be used here is 
shown in Fig.~\ref{ibm-tree}.
The branches in the tree are called \textit{tree branches};  all other
branches are called \textit{chords}.  
Each chord is associated with the one unique loop that is obtained when
adding the chord to the tree.  The orientation of a loop is
determined by the direction of its defining chord.  E.g., the
orientation of the loop 
pertaining to $L_1$ (large circle in Fig.~\ref{ibm-graph})
is anti-clockwise in Fig.~\ref{ibm-graph}.
\begin{figure}[t]
\centerline{\includegraphics[width=7cm]{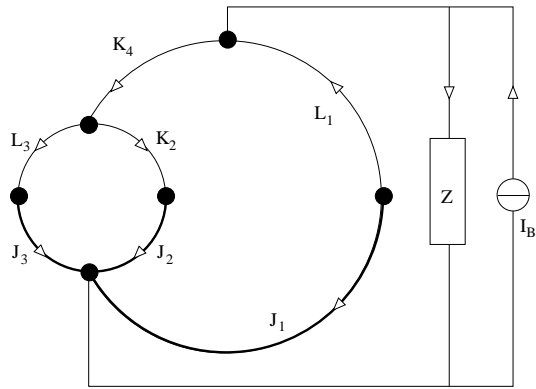}}
\caption{\label{ibm-graph}
The IBM qubit.  This is an example of a network graph with 6 nodes and 15 branches.  Each thick line represents a Josephson element, i.e.\ three branches in parallel, see Figure \ref{rsj}. Thin lines represent simple two-terminal elements, such as linear inductors (L, K), external impedances (Z), and current sources ($I_B$).}
\end{figure}
\begin{figure}[t]
\centerline{\includegraphics[width=3.5cm]{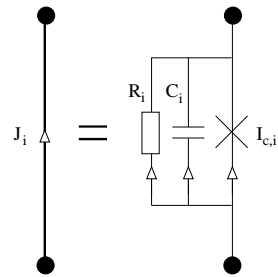}}
\caption{\label{rsj}
A Josephson subgraph (thick line) consists of three branches; a Josephson junction (cross), 
a shunt capacitor (C), a shunt resistor (R), and no extra nodes.}
\end{figure}
\begin{figure}[t]
\centerline{\includegraphics[width=4.5cm]{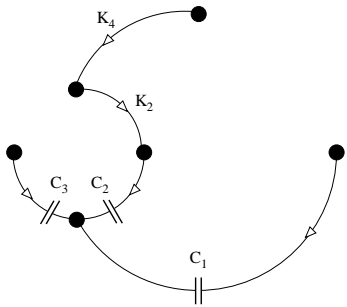}}
\caption{\label{ibm-tree}
A tree for the circuit shown in Figure \ref{ibm-graph}.  
A tree is a subgraph containing all nodes and no loop.  
Here, we choose a tree that contains all capacitors (C), some inductors (K), 
but no current sources ($I_B$) or external impedances (Z).}
\end{figure}

\item
Find the loop sub-matrices ${\bf F}_{CL}$, ${\bf F}_{CZ}$,
${\bf F}_{CB}$, ${\bf F}_{KL}$, ${\bf F}_{KZ}$, and ${\bf F}_{KB}$.
The loop sub-matrices have entries $+1$, $-1$, or $0$,
and hold the information about the important interconnections
in the circuit.  The matrix ${\bf F}_{XY}$ determines
which tree branches $X$ (either capacitors, $X=C$, or 
inductors $X=K$) are present in which loop defined by
the chords $Y$ (inductors, $Y=L$, external impedances $Y=Z$,
or current sources, $Y=B$).  In order to find, e.g.,
the loop sub-matrix ${\bf F}_{CL}$ for the IBM circuit
(Figs.~\ref{ibm-graph} and \ref{ibm-tree}), we have to identify
all loops obtained by adding a chord inductor ($L$).
Each column in ${\bf F}_{CL}$ corresponds to one such loop.
In our example, there are two chord inductors $L_1$ and
$L_3$;  the corresponding loops are the main superconducting
loop (large circle) and the control loop (small circle).
Each row in ${\bf F}_{CL}$ stands for one capacitor $C$;
therefore, in our example, ${\bf F}_{CL}$ is a 3 by 2 matrix.
The entries in each column of ${\bf F}_{CL}$ are $1$, $-1$,
or $0$, depending on whether the corresponding capacitor (row)
belongs to the corresponding loop (column) with the same ($-1$)
or opposite ($+1$) orientation or does not belong to the
loop at all ($0$).  E.g., for our example, [cf.\ Eq.~(\ref{FC-ibm-c})]
\[
  {\bf F}_{CL} = \left(\begin{array}{r r}
      1 &  0 \\
     -1 &  1 \\
      0 & -1
\end{array}\right).
\]
The first column says that the capacitor $C_1$ (part of $J_1$)
belongs to the large loop (in the opposite direction, thus $+1$), 
capacitor $C_2$ (part of $J_2$) belongs to the large loop 
(in the same direction, thus $-1$),
while capacitor $C_3$ (part of $J_3$) does not belong to the 
large loop at all.
Similarly, the second column of ${\bf F}_{CL}$ says which of
the capacitors are contained in the small loop.

\item
Use the inductances (self and mutual)
\begin{equation}
{\bf L}_{\rm t} =  \left(\begin{array}{c c}{\bf L}        & {\bf L}_{LK}\\
                                           {\bf L}_{LK}^T & {\bf L}_{K} 
                   \end{array}\right)
\end{equation}
and external impedances ${\bf Z}(\omega)$ to
calculate the matrices ${\bf M}_0$, ${\bf N}$, $\bar{\bf m}$, ${\bf S}$
using Eqs.~(\ref{M0}),(\ref{N}), (\ref{mbar}) and (\ref{IB});
for a single external impedance, also use Eqs.~(\ref{K-def})--(\ref{m})
to calculate the function $K(\omega)$, the coupling strength $\mu$ 
and the unit vector ${\bf m}$.
The block form of the inductance matrix ${\bf L}_{\rm t}$
originates from the distinction between tree (K) and chord (L) inductors;
${\bf L}$ is the chord inductance matrix (including 
chord-chord mutual inductances as its off-diagonal elements),
${\bf L}_K$ is the tree inductance matrix, and
${\bf L}_{LK}$ is the tree-chord mutual inductance matrix.
The Hamiltonian, Eqs.~(\ref{U})--(\ref{Hamiltonian-SB}),
together with the bath spectral density 
$J(\omega)\propto {\rm Im}K(\omega)$, Eq.~(\ref{JK}), 
represents the quantum theory 
of the system including the dissipative environment.
The form of this Hamiltonian, in particular the Equations
(\ref{M0})--(\ref{IB}) are the first main results of this paper.
The evolution of the density matrix $\rho$ of the superconducting
phases only is determined by the Bloch-Redfield equation (\ref{Redfield-equation})
with the Redfield tensor given by Eqs.~(\ref{RGamma})
and (\ref{Gp}), representing our second main result.

\item
Find the eigenstates and eigenenergies of the system Hamiltonian Eq.~(\ref{Hamiltonian-S})
and calculate the matrix elements of the superconducting phase operators $\bphi$.
In practice, this task is usually done numerically or using some approximation.
Typically, only a finite number of eigenstates is known.

\item
For two given quantum levels $|0\rangle$ and $|1\rangle$, the relaxation time $T_1$
and pure dephasing time $T_\phi$ can be found using Eqs.~(\ref{T1}) and (\ref{Tphi});
the decoherence time is then given by
\[
\frac{1}{T_2}=\frac{1}{2T_1}+\frac{1}{T_\phi}.
\]
The leakage rate $T_L^{-1}$ is given by Eq.~(\ref{Tleak}).
\end{enumerate}

We have carried out the above program for two cases;  
for the IBM qubit \cite{IBM} (Fig.~\ref{ibm-graph}) in Sec.~\ref{ibm-qubit}
and for the Delft qubit \cite{Mooij,Orlando} (Fig.~\ref{delft-graph}) in Sec.~\ref{delft-qubit}.
For the IBM qubit, matrix elements were calculated numerically;
the relaxation and decoherence times in the case of a current-biased circuit
are plotted in Fig.~\ref{ibm-T}.
For the Delft qubit, a semiclassical approach was taken, and earlier
results by van der Wal \textit{et al.} \cite{WWHM} for a symmetric
SQUID are correctly reproduced.
In addition to this, the effect of SQUID asymmetries--either in the
self inductance or in the critical currents of the two junctions--are
calculated in Sec.~\ref{delft-asym}.  It turns out that 
typical sample-to-sample fluctuations of the critical current
of about 10\% can lead to a sizable decoherence rate at zero
bias current.

\section{Classical network theory}
\label{classical}

The goal of this section is to derive a classical Hamiltonian for
an electrical circuit containing superconducting elements, such as
Josephson junctions.
An electric circuit will be represented by 
an oriented graph \cite{Peikari} ${\cal G}=({\cal N},{\cal B})$,
see Fig.~\ref{ibm-graph} for an example.

\subsection{Graph theory}
\label{graphs}
An oriented graph \footnote{Oriented graphs are sometimes referred to as directed graphs.
Strictly speaking, we are using \textit{multigraphs}, i.e., graphs
in which two nodes can be connected by more than one branch.} 
${\cal G}=({\cal N},{\cal B})$ consists of
$N$ nodes ${\cal N}=\{n_1,\ldots, n_N\}$ and $B$ branches
${\cal B}=\{b_1,\ldots, b_B\}$.
In circuit analysis, a branch $b_i=(n_{a(i)},n_{b(i)})$ represents 
a two-terminal element
(resistor, capacitor, inductor, current or voltage source, etc.), 
connecting its beginning node $n_{a(i)}$ to its ending node $n_{b(i)}$.
The degree of a node $n\in {\cal N}$ is the number of branches
containing $n$.
A loop in ${\cal G}$ is a subgraph of ${\cal G}$ in which all nodes have degree 2.
The number of disjoint connected subgraphs which, taken together, make up ${\cal G}$,
will be denoted $P$ and
the subgraphs ${\cal G}_i$, each having $N_i$ nodes and $B_i$ branches ($i=1,\ldots,P$),
where $\sum_{i=1}^P N_i=N$ and $\sum_{i=1}^P B_i=B$.
For each connected subgraph we choose a tree ${\cal T}_i$, i.e.\ a connected
subgraph of ${\cal G}_i$ which contains all its nodes and has no loops.
Note that ${\cal T}_i$ has exactly $N_i-1$ branches.
The $B_i-N_i+1$ branches that do not belong to the tree are called chords.
The tree of the graph ${\cal G}$ is the union of the trees of all its subgraphs,
${\cal T}_i$, containing $N-P$ branches.
A tree of the graph shown in Fig.~\ref{ibm-graph} is shown in Fig.~\ref{ibm-tree}.
The fundamental loops ${\cal F}_i$ of a subgraph ${\cal G}_i$ are defined as 
the set of loops in ${\cal G}_i$ which contain exactly one 
chord $f_i\in {\cal G}_i\backslash {\cal T}_i$.
We define the orientation of a fundamental loop via the orientation of its chord $f_i$.
Each connected subgraph ${\cal G}_i$ has $F_i = B_i - N_i + 1$ fundamental loops,
i.e.\ the graph has $F=\sum_{i=1}^P F_i=B-N+P$ fundamental loops (one for each chord).
A cutset of a connected graph is a set of a minimum number of branches
that, when deleted, divides the graph into two separate subgraphs.
A fundamental cutset of a graph with respect to a tree is a cutset that is
made up of one tree branch $c_i$ and a unique set of chords.
We denote the set of fundamental cutsets of ${\cal G}_i$ with respect to the
tree ${\cal T}_i$ with ${\cal C}_i$.  Each connected subgraph has $N_i - 1$
fundamental cutsets, therefore there are $N - P$ fundamental cutsets in total
(one for each tree branch).

We will use two characteristic matrices of the network graph,
the fundamental loop matrix ($i=1,\ldots F$; $j=1,\ldots, B$),
\begin{equation}
  {\bf F}^{(L)}_{ij} = \left\{\begin{array}{l l}
      1, & \mbox{if $b_j\in{\cal F}_i$ (same direction as $f_i$)},\\
     -1, & \mbox{if $b_j\in{\cal F}_i$ (direction opposite to $f_i$)},\\
      0, & \mbox{if $b_j\not\in{\cal F}_i$},
\end{array}\right. 
\end{equation}
and the fundamental cutset matrix ($i=1,\ldots, N+P$; $j=1,\ldots, B$),
\begin{equation}
  {\bf F}^{(C)}_{ij} = \left\{\begin{array}{l l}
      1, & \mbox{if $b_j\in{\cal C}_i$ (same direction as $c_i$)},\\
     -1, & \mbox{if $b_j\in{\cal C}_i$ (direction opposite to $c_i$)},\\
      0, & \mbox{if $b_j\not\in{\cal C}_i$}.
\end{array}\right. 
\end{equation}
By observing that cutsets always intersect loops in as many ingoing
as outgoing branches, one finds
\begin{equation}
  {\bf F}^{(L)} \left({\bf F}^{(C)} \right)^T  =  0, \label{BQ}
\end{equation}
By labeling the branches of the graph ${\cal G}$ 
such that the first $N-P$ branches belong to the tree ${\cal T}$, we obtain
\begin{equation}
  \label{QF}
  {\bf F}^{(C)} = \left( \openone \, |\, {\bf F}\right),
\end{equation}
where ${\bf F}$ is an $(N+P)\times(B-N-P)$ matrix.
Using Eq.~(\ref{BQ}), we find
\begin{equation}
  \label{BF}
  {\bf F}^{(L)} = \left(- {\bf F}^T \, |\, \openone \right).
\end{equation}

\subsection{Electric circuits}
\label{ecircuits}

The state of an electric circuit described by a network graph can be defined by the
branch currents ${\bf I}=(I_1,\ldots I_B)$, where $I_i$ denotes the
electric current flowing in branch $b_i$, and the branch 
voltages ${\bf V}=(V_1,\ldots V_B)$, where $V_i$ denotes the voltage drop across
the branch $b_i$.
The sign of $I_i$ is positive if a positive current flows from node 
$n_{a(i)}$ to $n_{b(i)}$ and negative if a positive current flows 
from node $n_{b(i)}$ to $n_{a(i)}$;  $V_i$ is positive if the 
electric potential is higher at node $n_{a(i)}$ than at node $n_{b(i)}$.

The conservation of electrical current, combined with the condition that no charge
can be accumulated at a node, implies Kirchhoff's current law,
\begin{equation}
  \label{Kirchhoff-Q}
  {\bf F}^{(C)}{\bf I} = 0.
\end{equation}
In a lumped circuit, energy conservation implies Kirchhoff's voltage law
in the form
\begin{equation}
  \label{Kirchhoff-B0}
  {\bf F}^{(L)}{\bf V} = 0.
\end{equation}
External magnetic fluxes ${\bf \Phi}=(\Phi_1,\ldots,\Phi_{B-N+P})$ 
threading the loops of 
the circuit represent a departure from the strict lumped circuit 
model;  if they are present, Faraday's law requires that
\begin{equation}
  \label{Kirchhoff-B}
  {\bf F}^{(L)}{\bf V} = {\bf \dot\Phi}.
\end{equation}
External fluxes have to be distinguished from the fluxes
associated with lumped circuit elements (e.g., inductors, see below).

We divide the branch currents and  voltages into a 
tree and a chord part,
\begin{eqnarray}
  \label{tree-chord-separation}
  {\bf I} &=& ({\bf I}_{\rm tr}, {\bf I}_{\rm ch}),\\
  {\bf V} &=& ({\bf V}_{\rm tr}, {\bf V}_{\rm ch}).
\end{eqnarray}
The $2B$ branch currents and voltages are not independent; the 
Kirchhoff laws Eqs.~(\ref{Kirchhoff-Q}) and (\ref{Kirchhoff-B})
together with  Eqs.~(\ref{QF}) and (\ref{BF}) yield the
following $B$ equations relating them,
\begin{eqnarray}
  {\bf F}  {\bf I}_{\rm ch} &=& - {\bf I}_{\rm tr},\label{IVct-1}\\
  {\bf F}^T{\bf V}_{\rm tr} &=&   {\bf V}_{\rm ch} - {\bf \dot\Phi}.\label{IVct-2}
\end{eqnarray}
As an example, the $N+P$ tree branch voltages ${\bf V}_{\rm tr}$
combined with the $B-N-P$ chord currents ${\bf I}_{\rm ch}$ completely
describe the state of a network, since all other currents and voltages
can be obtained from them via Eqs.~(\ref{IVct-1}) and (\ref{IVct-2}).
However, in the following, we will use a different subset of variables,
also making use of the
$B$ equations that are derived from the current-voltage
relations of the individual branch elements.

\subsection{Circuits containing superconducting elements}

For the purpose of analyzing electric circuits containing
Josephson junctions, we adopt the RSJ model for a Josephson junction,
i.e.\ a junction shunted by a capacitor and a resistor, see
Fig.~\ref{ibm-tree}.  We treat the Josephson junctions as
nonlinear inductors.  A (flux controlled) nonlinear 
inductor \cite{Peikari} is a two-terminal 
circuit element that follows a relation between the time-dependent
current $I(t)$ flowing through it and the voltage $V(t)$ across it 
of the form
\begin{equation}
  \label{nl-ind}
  I(t) = f(\Phi(t)),
\end{equation}
where $\dot\Phi(t)\equiv V (t)$ and $f$ is an arbitrary function.
For a linear inductor, $f(x)=x/L$, with $L$ the inductance.

We begin our analysis by choosing a tree containing all of the capacitors
in the network, no resistors
or external impedances, no current sources, and as few inductors as
possible (in particular, no Josephson junctions).
We assume here that the network does not 
contain any capacitor-only loops, which is realistic because in practice
any loop has a nonzero inductance.  A network is called proper if in addition
to this, it is possible to choose a tree without any inductors (i.e., if 
there are no inductor-only cutsets) \cite{Peikari}.
Again, it can be argued that this is realistic since there always are 
(at least small) capacitances between different parts of a network.
But we have avoided making the latter assumption here because it spares us
from describing the dynamics of small parasitic capacitances.  
We further assume
that each Josephson junction is shunted by a finite capacitance, so
that we are able to choose a tree without any Josephson junctions.
Finally, we assume for simplicity that the circuit does not contain any voltage
sources; however, voltage sources could easily be incorporated into our analysis.

We divide up the tree and chord currents and voltages further,
according to the various branch types,
\begin{eqnarray}
  \label{IVsplit}
  {\bf I}_{\rm tr}  =  ({\bf I}_C, {\bf I}_K), & & 
  {\bf I}_{\rm ch}  =  ({\bf I}_J, {\bf I}_L, {\bf I}_R, {\bf I}_Z, {\bf I}_B), \\
  {\bf V}_{\rm tr}  =  ({\bf V}_C, {\bf V}_K), & &
  {\bf V}_{\rm ch}  =  ({\bf V}_J, {\bf V}_L, {\bf V}_R, {\bf V}_Z, {\bf V}_B),
\quad\quad
\end{eqnarray}
where the tree current and voltage vectors contain a capacitor (C)
and tree inductor (K) part, whereas the chord current and voltage
vectors consist of parts for chord inductors, both non-linear (J) and linear (L),
shunt resistors (R) and other external impedances (Z), and bias current sources (B).
Accordingly, we write
\begin{eqnarray}
  \label{Fsplit}
  {\bf F} = \left(\begin{array}{c c c c c}
      {\bf F}_{CJ} & {\bf F}_{CL} & {\bf F}_{CR} & {\bf F}_{CZ} & {\bf F}_{CB} \\
      {\bf F}_{KJ} & {\bf F}_{KL} & {\bf F}_{KR} & {\bf F}_{KZ} & {\bf F}_{KB}
\end{array}\right).
\end{eqnarray}
The sub-matrices ${\bf F}_{XY}$ will be called \textit{loop sub-matrices}.
Note that since Josephson junctions are always shunted by a capacitor
as a tree branch, there are never any tree inductors in parallel with a
Josephson junction, ${\bf F}_{KJ}={\bf 0}$.  As a consequence,
a tree inductor is never in parallel with a shunt resistor, ${\bf F}_{KR}={\bf 0}$.

We then formally define the branch charges and fluxes ($X=C, K, J, L, R, Z, B$),
\begin{eqnarray}
  {\bf I}_X(t) &=& \dot{\bf Q}_X(t),    \label{charges}\\
  {\bf V}_X(t) &=& \dot{\bf \Phi}_X(t). \label{fluxes}
\end{eqnarray}
Using the second Josephson relation and Eq.~(\ref{fluxes}),
we identify the formal fluxes associated with the Josephson
junctions as the superconducting phase differences $\bphi$ across the junctions,
\begin{equation}
  \frac{{\bf \Phi}_J}{\Phi_0} =  \frac{\bphi}{2\pi} ,
\end{equation}
where $\Phi_0=h/2e$ is the superconducting flux quantum.
It will be assumed that at some initial time $t_0$
(which can be taken as $t_0\rightarrow -\infty$),
all charges and fluxes (including the external fluxes)
are zero, ${\bf Q}_X = 0$, ${\bf \Phi}_X = 0$ (including $\bphi =0$),
and ${\bf \Phi} =0$.

The current-voltage relations for the various types of branches are
\begin{eqnarray}
  {\bf I}_J  &=&  {\bf I}_{\rm c} \,\mbox{\boldmath $\sin$} \bphi ,\label{CVR-J}\\
  {\bf Q}_C  &=&  {\bf C}{\bf V}_C,\label{CVR-C}\\
  {\bf I}_L  &=&  \bar{\bf L}^{-1}{\bf \Phi}_L - {\bf L}^{-1}{\bf L}_{LK}\bar{\bf L}_K^{-1}{\bf \Phi}_K,\label{CVR-L}\\
  {\bf I}_K  &=&  \bar{\bf L}_K^{-1}{\bf \Phi}_K - {\bf L}_K^{-1}{\bf L}_{LK}^T\bar{\bf L}^{-1}{\bf \Phi}_L,\label{CVR-K}\\
  {\bf V}_R  &=&  {\bf R}{\bf I}_R\label{CVR-R},\\
  {\bf V}_Z(\omega)  &=&  {\bf Z}(\omega){\bf I}_Z(\omega)\label{CVR-Z},
\end{eqnarray}
where Eq.~(\ref{CVR-J}) is the first Josephson relation for the Josephson 
junctions (flux-controlled non-linear inductors), where the diagonal matrix 
${\bf I}_{\rm c}$ contains the critical currents $I_{{\rm c},i}$ of the junctions 
on its diagonal,
and $\mbox{\boldmath $\sin$}\bphi \equiv (\sin\varphi_1,\sin\varphi_2,\ldots,\sin\varphi_{N_J})$.
Eq.~(\ref{CVR-C}) describes
the (linear) capacitors (${\bf C}$ is the capacitance matrix), Eqs.~(\ref{CVR-L}) and (\ref{CVR-K})
the linear inductors, see Eqs.~(\ref{Lbar}) and (\ref{LKbar}) below.  
The junction shunt resistors are described by 
Eq.~(\ref{CVR-R}) where $R$ is the (diagonal and real) shunt resistance matrix.
The external impedances are described by the relation
Eq.~(\ref{CVR-Z}) between the Fourier transforms of the
current and voltage,  where ${\bf Z}(\omega)$ is the impedance matrix.
The external impedances can also defined in the time domain,
\begin{equation}
  \label{CVR-Z-time}
  {\bf V}_Z(t)=\int_{-\infty}^t {\bf Z}(t-\tau){\bf I}_Z(\tau) d\tau \equiv ({\bf Z}*{\bf I}_Z)(t),
\end{equation}
where the convolution is defined as
\begin{equation}
  \label{convolution}
  ({\bf f} * {\bf g}) (t) = \int_{-\infty}^t {\bf f}(t - \tau) {\bf g}(\tau) d\tau .
\end{equation}
Causality allows the response function to be nonzero only for 
positive times, ${\bf Z}(t)=0$ for $t<0$.  
In frequency space, the replacement $\omega\rightarrow \omega + i\epsilon$
with $\epsilon>0$ guarantees convergence of the Fourier transform
\footnote{We choose the Fourier transform such that it yields the impedance
$Z(\omega)=+i\omega L$ for an inductor (inductance $L$).}
\begin{equation}
  \label{Z-FT}
  {\bf Z}(\omega) = \int_{-\infty}^\infty {\bf Z}(t)e^{i\omega t}dt
                  = \int_0^\infty {\bf Z}(t)e^{i\omega t}dt .
\end{equation}
In order to obtain Eq.~(\ref{CVR-L}) for the inductors, we write
\begin{equation}
  \label{inductance-1}
   \left(\begin{array}{c}{\bf \Phi}_L\\ {\bf \Phi}_K\end{array}\right)
  =\left(\begin{array}{l l}{\bf L}        & {\bf L}_{LK}\\
                           {\bf L}_{LK}^T & {\bf L}_{K} \end{array}\right)
   \left(\begin{array}{c}{\bf I}_L\\ {\bf I}_K\end{array}\right)
  \equiv {\bf L}_{\rm t} \left(\begin{array}{c}{\bf I}_L\\ {\bf I}_K\end{array}\right),
\end{equation}
where ${\bf L}$ and ${\bf L}_K$ are the self inductances of the chord and tree branch
inductors, resp., off-diagonal elements describing the mutual inductances among
chord inductors and tree inductors separately, and  ${\bf L}_{LK}$ is the
mutual inductance matrix between tree and chord inductors.
Since the total inductance matrix is symmetric and positive, i.e.\
${\bf v}^T {\bf L}_{\rm t} {\bf v}>0$ for all real vectors ${\bf v}$,
its inverse exists, and we find
\begin{eqnarray}
   \left(\begin{array}{c}{\bf I}_L\\ {\bf I}_K\end{array}\right)
  &=&\left(\begin{array}{c c}\bar{\bf L} ^{-1}    & -{\bf L}^{-1}{\bf L}_{LK}\bar{\bf L}_K^{-1}\\
                   -{\bf L}_K^{-1}{\bf L}_{LK}^T\bar{\bf L}^{-1} & \bar{\bf L}_{K} ^{-1} \end{array}\right)
   \left(\begin{array}{c}{\bf \Phi}_L\\ {\bf \Phi}_K\end{array}\right)\nonumber\\
  &\equiv&{\bf L}_{\rm t}^{-1}\left(\begin{array}{c}{\bf \Phi}_L\\ {\bf \Phi}_K\end{array}\right)
\label{inductance-2}
\end{eqnarray}
with the definitions
\begin{eqnarray}
  \bar{\bf L}        &=& {\bf L}   - {\bf L}_{LK}   {\bf L}_K^{-1} {\bf L}_{LK}^T,\label{Lbar}\\
  \bar{\bf L}_{K}    &=& {\bf L}_K - {\bf L}_{LK}^T {\bf L}  ^{-1} {\bf L}_{LK}.\label{LKbar}
\end{eqnarray}
Note that the matrices ${\bf L}$ and ${\bf L}_K$, being diagonal sub-matrices of a
symmetric and positive matrix, are also symmetric and positive and thus their 
inverses exist.
The operators $\bar{\bf L}$ and $\bar{\bf L}_{K}$ as defined in 
Eqs.~(\ref{Lbar}) and (\ref{LKbar}) are invertible since ${\bf L}_{\rm t}^{-1}$ exists.
Moreover, since the inverse of the total inductance matrix, 
see Eq.~(\ref{inductance-2}), is symmetric and positive, its diagonal sub-matrices
are symmetric and positive, and thus $\bar{\bf L}, \bar{\bf L}_{K} >0 $.

\subsection{Equations of motion}
\label{eq-mot}

In order to derive a Lagrangian for an electric circuit, we have
to single out among the charges and fluxes a complete set of unconstrained 
degrees of freedom, such that each assignment of values to those charges
and fluxes and their first time derivatives 
represents a possible dynamical state of the system.
Using Eqs.~(\ref{Fsplit}--\ref{fluxes}), (\ref{CVR-J}--\ref{CVR-Z}),
(\ref{inductance-1}), and (\ref{inductance-2}),  the time evolution 
of the charges and fluxes can be expressed as the following set of first-order
integro-differential equations
\begin{eqnarray}
  \frac{\Phi_0}{2\pi}\dot\bphi  &=& {\bf V}_{J}
                  = {\bf F}_{CJ}^T {\bf C}^{-1}{\bf Q}_C, \label{eqm1-J}\\
  \dot{\bf Q}_{C} &=& {\bf I}_{C}
                  = -{\bf F}_{CJ} {\bf I}_{\rm c}\, \mbox{\boldmath $\sin$} \bphi
                  -{\bf F}_{CR} {\bf R}^{-1}\dot{\bf \Phi}_{R} \nonumber\\ 
&&                -{\bf F}_{CL} \left(\bar{\bf L}^{-1}{\bf \Phi}_L
                  -{\bf L}^{-1}{\bf L}_{LK}\bar{\bf L}_K^{-1}{\bf \Phi}_K\right) \nonumber\\ &&
                  -{\bf F}_{CZ} {\bf L}_Z^{-1}*{\bf \Phi}_Z -{\bf F}_{CB}{\bf I}_B, \label{eqm1-C}\\
  \dot{\bf \Phi}_{L} &=& {\bf V}_{\rm L}
                = {\bf F}_{CL}^T {\bf C}^{-1}{\bf Q}_C + {\bf F}_{KL}^T \dot{\bf \Phi}_K
                  +\dot{\bf \Phi}_{x}, \label{eqm1-L}\\
  \dot{\bf \Phi}_{R} &=& {\bf V}_{\rm R}
                = {\bf F}_{CR}^T {\bf C}^{-1}{\bf Q}_C, \label{eqm1-R}\\
  \dot{\bf \Phi}_{Z} &=& {\bf V}_{\rm Z}
                = {\bf F}_{CZ}^T {\bf C}^{-1}{\bf Q}_C + {\bf F}_{KZ}^T \dot{\bf \Phi}_K, \label{eqm1-Z}\\
  {\bf \Phi}_{K}     &=& -{\bf L}_K \bar{\bf F}_{KL}\bar{\bf L}^{-1} {\bf \Phi}_L
                         +{\bf L}_K \bar{\bf F}_{KL}{\bf L}^{-1} 
                          {\bf L}_{LK}\bar{\bf L}_K^{-1} {\bf \Phi}_K \nonumber\\  &&  
-{\bf L}_K {\bf F}_{KZ}{\bf L}_Z^{-1} * {\bf \Phi}_Z 
                         -{\bf L}_K {\bf F}_{KB} {\bf I}_B,\quad\label{eqm1-K}
\end{eqnarray}
where ${\bf L}_Z(\omega)\equiv{\bf Z(\omega)}/i\omega$, and where the convolution is given by
Eq.~(\ref{convolution}).
In the equations for the chord variables Eqs.~(\ref{eqm1-J}), (\ref{eqm1-L}), (\ref{eqm1-R}),
and (\ref{eqm1-Z}), we have assumed that only the loops closed by a chord inductor (L)
are threaded by an external flux, ${\bf\Phi}=(0,{\bf \Phi}_x,0,0,0)$.
In order to obtain Eq.~(\ref{eqm1-K}), we have first used Eq.~(\ref{inductance-1}), then
Eqs.~(\ref{IVct-1}) and (\ref{CVR-Z}), and finally Eq.~(\ref{inductance-2}).
We can eliminate ${\bf \Phi}_{K}$ by solving Eq.~(\ref{eqm1-K}),
\begin{equation}
  \label{sol-L}
  {\bf \Phi}_{K} = -\tilde{\bf L}_K\left(\bar{\bf F}_{KL}\bar{\bf L}^{-1}{\bf \Phi}_L+{\bf F}_{KZ}{\bf L}_Z^{-1}*{\bf \Phi}_Z+{\bf F}_{KB}{\bf I}_B\right),
\end{equation}
with the definitions
\begin{eqnarray}
  \tilde{\bf L}_K &=& \left(\openone_K -{\bf L}_K\bar{\bf F}_{KL}{\bf L}^{-1}{\bf L}_{LK}\bar{\bf L}_K^{-1}\right)^{-1} {\bf L}_K , \label{LKtilde}\\
  \bar{\bf F}_{KL}&=& {\bf F}_{KL} -{\bf L}_K^{-1}{\bf L}_{LK}^T \label{FKLtilde}.
\end{eqnarray}

Further knowledge of the structure of ${\bf F}$ can be derived from the fact that
Josephson junctions are always assumed to be RC-shunted, see Fig.~\ref{rsj}.
If we label the tree branches such that the first $N_J\le N_C$ capacitances are the ones
shunting the Josephson junctions ($N_C$=number of capacitances, $N_J$=number of
Josephson junctions) then we find
\begin{eqnarray}
  \label{F-rsj-mixed}
  {\bf F}_{CJ} &=& {\bf F}_{CR} = \left(\begin{array}{c}\openone _{N_J}\\ 0_{N_C-N_J}\end{array}\right),\\
  {\bf Q}_C    &=& \left(\begin{array}{c} {\bf Q}_{CJ}\\ {\bf Q}_{\bar{C}}\end{array}\right),
\end{eqnarray}
where $\bar{C}$ denotes the capacitors which are not parallel shunts of a Josephson junction.
In general, the charges of these additional capacitors represent independent degrees of
freedom in addition to the shunt capacitor charges ${\bf Q}_{CJ} = \Phi_0 {\bf C} \dot\bphi / 2\pi$.
But from this point onward, we will study the case where there are no capacitors except the
Josephson junction shunt capacitors, $N_C=N_J$.
However, the resulting equation of motion (\ref{eq-motion-1}) with
the definitions Eqs.~(\ref{M0})--(\ref{IB}) still allows us to describe pure capacitors
by treating them as Josephson elements with zero critical current $I_c$ and 
infinite shunt resistance $R$.
With this simplification,
\begin{equation}
  \label{F-rsj-s}
  {\bf F}_{CJ} = {\bf F}_{CR} = \openone ,
\end{equation}
and the $\bphi$ and $\dot\bphi$ can be chosen as the $2N_J$
generalized coordinates and velocities that satisfy the equation of motion
\begin{eqnarray}
  {\bf C}\ddot\bphi &=&  - {\bf L}_J^{-1} \mbox{\boldmath $\sin$} \bphi
                         - {\bf R}^{-1} \dot\bphi \label{eq-motion-0}\\
                      && -  \frac{2\pi}{\Phi_0}
                            \left({\bf F}_{CL} \tilde{\bf L}_L^{-1}{\bf \Phi}_L
                           + \bar{\bf F}_{CZ}{\bf L}_Z^{-1}*{\bf \Phi}_Z
                           + \bar{\bf F}_{CB}{\bf I}_B \right),\nonumber
\end{eqnarray}
where we have used Eqs.~(\ref{eqm1-J}), (\ref{eqm1-C}), and (\ref{sol-L}), 
and introduced ${\bf L}_J^{-1} = 2\pi {\bf I}_{\rm c}/\Phi_0$, and ($Y=Z,B$)
\begin{eqnarray}
  \tilde{\bf L}_L^{-1}  &=&  \left(\openone_L +{\bf L}^{-1}{\bf L}_{LK}\bar{\bf L}_K^{-1}\tilde{\bf L}_K\bar{\bf F}_{KL}\right)\bar{\bf L}^{-1}, \label{LLtilde}\\
\bar{\bf F}_{CY} &=& {\bf F}_{CY}+ {\bf F}_{CL}{\bf L}^{-1}{\bf L}_{LK}\bar{\bf L}^{-1}_{K}\tilde{\bf L}_K {\bf F}_{KY}.\label{FCYtilde}
\end{eqnarray}
The remaining state variables obey the following linear relations,
\begin{eqnarray}
  {\bf L}_{LL} \bar{\bf L}^{-1}   \dot{\bf\Phi}_L + {\bf L}_{LZ} {\bf L}_Z^{-1}*\dot{\bf\Phi}_Z 
  &=& {\bf a}_L(\dot\bphi),\label{ZL-1}\\
  {\bf L}_{ZL} \bar{\bf L}^{-1}\dot{\bf\Phi}_L + {\bf L}_{ZZ}  {\bf L}_Z^{-1}*\dot{\bf\Phi}_Z
  &=& {\bf a}_Z(\dot\bphi),\label{ZL-2}
\end{eqnarray}
where we have introduced
\begin{eqnarray}
  {\bf L}_{LL} &=& \bar{\bf L} + {\bf F}_{KL}^T \tilde{\bf L}_K \bar{\bf F}_{KL}, \label{LLL}\\
  {\bf L}_{ZZ} &=& {\bf L}_Z  + {\bf F}_{KZ}^T \tilde{\bf L}_K {\bf F}_{KZ},\label{LZZ}\\
  {\bf L}_{LZ} &=& {\bf F}_{KL}^T \tilde{\bf L}_K {\bf F}_{KZ} ,\label{LLZ}\\
  {\bf L}_{ZL} &=& {\bf F}_{KZ}^T \tilde{\bf L}_K \bar{\bf F}_{KL} ,\label{LZL}\\
  {\bf a}_L(\dot\bphi)    &=& \frac{\Phi_0}{2\pi}{\bf F}_{CL}^T \dot\bphi 
                    + \dot{\bf \Phi}_x
                    - {\bf F}_{KL}^T \tilde{\bf L}_K {\bf F}_{KB} \dot{\bf I}_B,\label{aL}\\
  {\bf a}_Z(\dot\bphi)    &=& \frac{\Phi_0}{2\pi}{\bf F}_{CZ}^T \dot\bphi 
     - {\bf F}_{KZ}^T \tilde{\bf L}_K {\bf F}_{KB} \dot{\bf I}_B.\label{aZ}
\end{eqnarray}
Note that in the absence of dissipation, ${\bf L}_Z^{-1}\rightarrow{\bf 0}$, 
Eqs.~(\ref{ZL-1}) and (\ref{ZL-2}) are holonomic constraints for the
variables $\dot{\bf\Phi}_L$, since 
Eqs.~(\ref{ZL-1}) and (\ref{ZL-2}) can be integrated.
If ${\bf L}_{LL}$, ${\bf L}_{ZZ}$, and
\begin{eqnarray}
  \bar{\bf L}_{L} &=& {\bf L}_{LL} - {\bf L}_{LZ}{\bf L}_{ZZ}^{-1}{\bf L}_{ZL},\label{LLbar}\\
  \bar{\bf L}_{Z} &=& {\bf L}_{ZZ} - {\bf L}_{ZL}{\bf L}_{LL}^{-1}{\bf L}_{LZ},\label{LZbar}
\end{eqnarray}
are regular matrices, the solution to Eqs.~(\ref{ZL-1}) and (\ref{ZL-2}) is given by
\begin{eqnarray}
  \dot{\bf\Phi}_L &=& \bar{\bf  L}\bar{\bf L}_{L}^{-1}\left({\bf a}_L(\dot\bphi) - {\bf L}_{LZ} {\bf L}_{ZZ}^{-1}*{\bf a}_Z(\dot\bphi)\right),\label{sol-I}\\
  \dot{\bf\Phi}_Z &=& {\bf  L}_Z\bar{\bf L}_{Z}^{-1}*\left({\bf a}_Z(\dot\bphi) - {\bf L}_{ZL} {\bf L}_{LL}^{-1}{\bf a}_L(\dot\bphi)\right). \label{sol-Z}
\end{eqnarray}
Note that in the limit of large external impedances, ${\bf L}_Z\rightarrow{\bf 0}$,
the regularity conditions for ${\bf L}_{LL}$, ${\bf L}_{ZZ}$, $\bar{\bf L}_{L}$,
and $\bar{\bf L}_{Z}$ all collapse to the condition that ${\bf L}_{LL}$ be regular.
The latter always holds in the absence of mutual inductances between tree and chord
inductors, since in this case $\bar{\bf F}_{KL}={\bf F}_{KL}$ and thus ${\bf L}_{LL}$
is symmetric and positive, so that its inverse exists.
Integrating Eqs.~(\ref{sol-I}) and (\ref{sol-Z}) from $t_0$ to $t$, 
using the initial condition (all charges and fluxes equal to zero), 
and substituting the solutions into Eq.~(\ref{eq-motion-0}), we arrive at
the classical equation of motion for the superconducting phases $\bphi$,
\begin{widetext}
\begin{equation}
    {\bf C}\ddot\bphi =    - {\bf L}_{J}^{-1}{\mbox{\boldmath $\sin$}}\bphi
                           - {\bf R}^{-1}\dot\bphi 
                           - {\bf M}_0\bphi
                           - {\bf M}_d * \bphi 
                           - \frac{2\pi}{\Phi_0}{\bf N}{\bf \Phi}_x
                           - \frac{2\pi}{\Phi_0}{\bf S}{\bf I}_{B}, \label{eq-motion-1}
\end{equation}
\end{widetext}
with
\begin{eqnarray}
  {\bf M}_0   &=& {\bf F}_{CL} \tilde{\bf L}_L^{-1}\bar{\bf  L}{\bf L}_{LL}^{-1}{\bf F}_{CL}^T,
                               \label{M0}\\
  {\bf N}     &=& {\bf F}_{CL} \tilde{\bf L}_L^{-1}\bar{\bf  L}{\bf L}_{LL}^{-1},\label{N}\\
  {\bf M}_d(\omega) &=&  \bar{\bf m}\bar{\bf L}_Z^{-1}(\omega)\bar{\bf m}^T, \label{Md}\\ 
  \bar{\bf m} &=& {\bf F}_{CZ} - {\bf F}_{CL} ({\bf L}_{LL}^{-1})^T \bar{\bf F}_{KL}^T \tilde{\bf L}_K^T {\bf F}_{KZ}, \label{mbar}\\
  {\bf S} &=& {\bf F}_{CB} - {\bf F}_{CL} ({\bf L}_{LL}^{-1})^T \bar{\bf F}_{KL}^T \tilde{\bf L}_K^T {\bf F}_{KB}. \label{IB}
\end{eqnarray}
Although the expression (\ref{M0}) for the matrix ${\bf M}_0$ is not manifestly
symmetric, we show in Appendix \ref{M0-sym} that it is indeed symmetric, i.e.\ 
${\bf M}_0^T={\bf M}_0$.  This property of ${\bf M}_0$ allows us to write the term ${\bf M}_0\bphi$
in the equations of motion (\ref{eq-motion-1}) as the gradient of a potential,
see Eq.~(\ref{U}) below.
The matrices ${\bf M}_d(\omega)$ and ${\bf R}$ contain all the dissipative dynamics of $\bphi$;
if all external impedances (shunt resistors) 
are removed, then ${\bf L}_Z^{-1}\rightarrow {\bf 0}$ and thus
${\bf M}_d(\omega)\rightarrow {\bf 0}$ (${\bf R}^{-1}\rightarrow 0$).
A proof of the symmetry of the dissipation matrix, ${\bf M}_d={\bf M}_d^T$, and a derivation 
of the representation in Eqs.~(\ref{Md}) and (\ref{mbar}) can be found in Appendix \ref{Md-sym}.

Note that the coupling matrix ${\bf S}$ to an external bias current ${\bf I}_B$ can
be obtained from $\bar{\bf m}$ by replacing $Z$ by $B$.  Physically, this means that
the external impedances $Z$ can be thought of as fluctuating external currents;  in
particular, if a bias current is shunted in parallel to an impedance,
${\bf F}_{XZ}=\pm {\bf F}_{XB}$ ($X=C,K$) 
then we find ${\bf S}=\pm\bar{\bf m}$.
In deriving the equation of motion (\ref{eq-motion-1}), 
we have assumed that the external magnetic fluxes and bias currents become time-independent
after they have been switched on in the past, $\dot{\bf \Phi}_x\rightarrow 0$,
$\dot{\bf I}_B\rightarrow 0$ ($t>t_0$).
In the absence of mutual inductances between the tree and chord inductors, ${\bf L}_{KL}={\bf 0}$, 
Eqs.~(\ref{M0})--(\ref{IB}) become somewhat simpler,
\begin{eqnarray}
  {\bf M}_0    &=& {\bf F}_{CL} {\bf L}_{LL}^{-1}{\bf F}_{CL}^T,\label{M0-s}\\
  {\bf N}      &=& {\bf F}_{CL} {\bf L}_{LL}^{-1},\label{N-s}\\
  \bar{\bf m}  &=& {\bf F}_{CZ} - {\bf F}_{CL} {\bf L}_{LL}^{-1} {\bf F}_{KL}^T {\bf L}_K {\bf F}_{KZ}, \label{mbar-s}\\
  {\bf S}  &=& {\bf F}_{CB} - {\bf F}_{CL} {\bf L}_{LL}^{-1} {\bf F}_{KL}^T {\bf L}_K {\bf F}_{KB},\label{IB-s}\\
  {\bf L}_{LL} &=& {\bf L}+{\bf F}_{KL}^T {\bf L}_K {\bf F}_{KL} = {\bf L}_{LL}^T.\label{LLL-s}
\end{eqnarray}

It should be noted here that from now on, the shunt resistors ${\bf R}$ can be treated
as external impedances by setting
${\bf M}_d' =  {\bf M}_d + i\omega {\bf R}^{-1}$;  the only reason for
treating the shunt resistors separately is that more is known about the 
possible arrangement
of the shunt resistors in the circuit.  We will mostly concentrate on external impedances
in our examples and neglect the shunt resistors, because in our examples $R\gg Z$.
If, in turn, the external impedances are pure resistors, i.e.\
${\bf Z}(\omega)$ is real and frequency independent, then they can be described as
corrections to ${\bf R}$, i.e.\ ${\bf R}' = {\bf R} + {\bf Z}$.

A few important remarks about the form of the matrix ${\bf M}_d$ are in order.
(i) We know that ${\bf M}_d(t)$ is real, causal (i.e., ${\bf M}_d(t)=0$ for $t<0$),
and symmetric ${\bf M}_d={\bf M}_d^T$ (Appendix \ref{Md-sym}).
A dissipative term in the equations of motion with these properties can be modeled
using the Caldeira-Leggett formalism \cite{CaldeiraLeggett}.
(ii) In the lowest-order Born approximation, i.e.\ perturbation
theory in the equation of motion in the small parameters $Z_i^{-1}$ (see below),
the contributions to ${\bf M}_d$ from different external impedances are additive,
in the sense that
one can calculate ${\bf M}_d$ for each impedance $Z_i$ separately,
while $Z_{j\neq i}\rightarrow\infty$, and then add the contributions
in order to obtain the full coupling Hamiltonian (see Eq.~(\ref{Hamiltonian-SB}) below).
In the same manner, the decoherence rates due to different impedances will be
additive in the lowest-order Born approximation.
An exact statement (independent of the Born approximation) can be made if 
$\bar{\bf L}_Z^{-1}$ can be written
as a sum in which every term contains only one of the impedances $Z_i$,
since in this case ${\bf M}_d=\sum_{i=1}^{N_Z}{\bf M}_{d,i}$ where $N_Z$ denotes 
the number of external impedances and ${\bf M}_{d,i}(\omega)$ describes the effect of $Z_i$.
From now on, we will study the case of a single external impedance, bearing in mind that 
in lowest-order perturbation theory the results obtained in this way can easily be 
used to describe the dynamics of a system coupled to several external impedances.
(iii) In the case of a single impedance, ${\bf M}_d (\omega)$ has the form,
\begin{eqnarray}
   {\bf M}_d (\omega) &=& \mu K(\omega) {\bf m}{\bf m}^T,\label{Md-simple}\\
   K(\omega)          &=& \bar{\bf L}_Z^{-1}(\omega),\label{K-def}\\
   \mu                &=& |\bar{\bf m}|^2,\label{mu}\\
   {\bf m}            &=& \bar{\bf m}/\sqrt{\mu}=\bar{\bf m}/|\bar{\bf m}|,\label{m}
\end{eqnarray}
where $K(t)$ is a scalar real function, ${\bf m}$ is the normalized
vector parallel to $\bar{\bf m}$, and $\sqrt{\mu}$ is the length 
of the vector $\bar{\bf m}$
($\mu$ is the eigenvalue of the rank 1 matrix $\bar{\bf m}\bar{\bf m}^T$).

The dissipation free (${\bf R}, {\bf Z}\rightarrow \infty$, ${\bf M}_d=0$) 
part of the classical equation of motion
Eq.~(\ref{eq-motion-1}) can be derived from the Lagrangian
\begin{eqnarray}
{\cal L}_0 &=&
\left(\frac{\Phi_0}{2\pi}\right)^2 \left(\frac{1}{2}\dot{\bphi}^T{\bf C}\dot{\bphi} - U(\bphi)\right),\\
U(\bphi) &=& -\sum_i L_{J;i}^{-1}\cos\varphi_i \nonumber\\ && 
             + \frac{1}{2}\bphi^T {\bf M}_0 \bphi 
             + \frac{2\pi}{\Phi_0}\bphi^T \left({\bf N} {\bf \Phi}_x 
                                                  + {\bf S}{\bf I}_B \right),\label{U}
\end{eqnarray}
or, equivalently, from the Hamiltonian
\begin{eqnarray}
{\cal H}_S = \frac{1}{2}{\bf Q}_C^T {\bf C}^{-1}{\bf Q}_C 
            + \left(\frac{\Phi_0}{2\pi}\right)^2  U(\bphi),   \label{Hamiltonian-S}
\end{eqnarray}
where the canonical momenta corresponding to the flux variables $\Phi_0\bphi/2\pi$
are the capacitor charges
\[
  \frac{2\pi}{\Phi_0}\frac{\partial {\cal L}_0}{\partial\dot\bphi} 
       = \frac{\Phi_0}{2\pi}{\bf C}\dot\bphi 
       = {\bf C}{\bf V}_C \equiv {\bf Q}_C.
\]

\section{Canonical quantization of  ${\cal H}_S$ and system-bath model}
\label{quantum}

In this Section, we quantize the classical theory for a superconducting circuit
that was derived in the previous Section.
The conjugate flux and charge variables $\bphi$ and ${\bf Q}_C$ now have to be
understood as operators with the commutation relations
\begin{equation}
  \label{commutator}
  \left[\frac{\Phi_0}{2\pi}\varphi_i , Q_{C;j} \right] = i\hbar\delta_{ij} .
\end{equation}
In order to include the dissipative dynamics of the classical equation of motion,
Eq.~(\ref{eq-motion-1}),
in our quantum description, we follow Caldeira and Leggett \cite{CaldeiraLeggett},
and introduce a bath (reservoir) of harmonic oscillators describing the
degrees of freedom of the external impedances.
We will restrict ourselves to the case of a single external impedance coupled to
the circuit (this is sufficient to describe the general case in the lowest-order 
Born approximation, see Sec.~\ref{classical}). 
For the Hamiltonian of the circuit including the external impedance, we write
\begin{eqnarray}
{\cal H}   &=& {\cal H}_S + {\cal H}_B + {\cal H}_{SB},\label{Hamiltonian}\\
{\cal H}_B &=& \frac{1}{2}\sum_\alpha\left(\frac{p_\alpha^2}{m_\alpha}+m_\alpha \omega_\alpha^2 x_\alpha^2\right),\label{Hamiltonian-B}\\
{\cal H}_{SB} &=&  {\bf m}\cdot\bphi \sum_\alpha c_\alpha x_\alpha  + \Delta U(\bphi),\label{Hamiltonian-SB}
\end{eqnarray}
where ${\cal H}_S$ is the quantized Hamiltonian Eq.~(\ref{Hamiltonian-S}), 
derived in Sec.~\ref{classical}, ${\cal H}_B$ is the Hamiltonian describing 
a bath of harmonic oscillators with (fictitious) position and momentum 
operators $x_\alpha$ and $p_\alpha$ with $[x_\alpha, p_\beta]=i\hbar\delta_{\alpha\beta}$, 
masses $m_\alpha$, and oscillator 
frequencies $\omega_\alpha$.  Finally, ${\cal H}_{SB}$ describes the coupling 
between the system and bath degrees of freedom, $\bphi$ and $x_\alpha$, where 
$c_\alpha$ is a coupling parameter and ${\bf m}$ is defined in Eqs.~(\ref{mbar}) and (\ref{m}).  
The term $\Delta U(\bphi)=({\bf m}\cdot \bphi)^2\sum_\alpha c_\alpha^2/2m_\alpha\omega_\alpha^2$ 
compensates the energy renormalization caused by the system-bath interaction 
(first term) \cite{CaldeiraLeggett}.
It ensures that, for a fixed value of $\bphi$,
\begin{equation}
\min_{\{x_\alpha\}} \left[U(\bphi) + {\cal H}_B(\{x_\alpha\}) + {\cal H}_{SB}(\bphi,\{x_\alpha\})\right]
  = U(\bphi),
\end{equation}
or, equivalently, that for all $\bphi$
\begin{equation}
\min_{\{x_\alpha\}} \left[{\cal H}_B(\{x_\alpha\}) + {\cal H}_{SB}(\bphi,\{x_\alpha\})\right]
  = 0.
\end{equation}
The term  $\Delta U(\bphi)$ will not be relevant for the Redfield theory to be derived below.

In Eq.~(\ref{Hamiltonian-SB}), we have already anticipated the form of the 
system-bath interaction.  In order to verify this and to determine the spectral
density of the bath (the masses, frequencies, and coupling constants will only 
enter through this quantity, see below), we derive the classical 
equations of motion from the Hamiltonian Eq.~(\ref{Hamiltonian}) in the
Fourier representation.  The equations of motion for the bath variables are
\begin{equation}
  -\omega^2 m_\alpha x_\alpha = -m_\alpha \omega_\alpha^2 x_\alpha -c_\alpha {\bf m}\cdot \bphi.
\end{equation}
Solving for $x_\alpha$, we obtain
\begin{equation}
\label{sol-x}
  x_\alpha = c_\alpha \frac{{\bf m}\cdot\bphi }{m_\alpha(\omega^2 - \omega_\alpha^2)}.
\end{equation}
The equation of motion for $\bphi$ is
\begin{equation}
  -\omega^2 {\bf C}\bphi  = -\frac{\partial U}{\partial \bphi}
      - \left(\frac{2\pi}{\Phi_0}\right)^2 {\bf m} \sum_\alpha c_\alpha x_\alpha .
\end{equation}
Using Eq.~(\ref{sol-x}), we find
\begin{equation}
\label{sol-phi}
  -\omega^2 {\bf C}\bphi = - \frac{\partial U}{\partial \bphi}
     -\!\left(\frac{2\pi}{\Phi_0}\right)^2 \!\!\! {\bf m} ({\bf m}\cdot\bphi)
    \!\sum_\alpha\frac{c_\alpha^2}{m_\alpha (\omega^2 \!-\!\omega_\alpha^2)}.\quad
\end{equation}
Comparing Eq.~(\ref{sol-phi}) to the Fourier transform of Eq.~(\ref{eq-motion-1}),
and using the decomposition Eqs.~(\ref{Md-simple}) 
we obtain the expression
\begin{equation}
\label{K}
  K(\omega ) = \frac{1}{\mu}
   \left(\frac{2\pi}{\Phi_0}\right)^2\sum_\alpha\frac{c_\alpha^2}{m_\alpha (\omega^2-\omega_\alpha^2)}.
\end{equation}
The spectral density of a bath of harmonic oscillators is defined as \cite{CaldeiraLeggett}
\begin{equation}
\label{J}
  J(\omega) = \frac{\pi}{2}\sum_\alpha\frac{c_\alpha^2}{m_\alpha \omega_\alpha}\delta(\omega-\omega_\alpha);
\end{equation}
combining Eqs.~(\ref{K}) and (\ref{J}), we arrive at
\begin{equation}
  K(\omega) = \frac{1}{\mu}\left(\frac{2\pi}{\Phi_0}\right)^2 
               \frac{2}{\pi}\int_0^\infty d\omega' \frac{\omega' J(\omega')}{\omega^2-\omega'^2} .
\end{equation}
We now use the replacement $K(\omega)\rightarrow K(\omega +i\epsilon)$,
since $K(\omega)$ is a function of the external impedance $Z(\omega)$, see Eq.~(\ref{Z-FT}),
\[
  \frac{1}{\omega-\omega'} = \lim_{\epsilon\rightarrow 0}\frac{1}{\omega-\omega'+i\epsilon} = P\frac{1}{\omega-\omega'}-i\pi\delta(\omega'-\omega),
\]
and obtain
\begin{equation}
  K(\omega)
  = \frac{1}{\mu}\left(\frac{2\pi}{\Phi_0}\right)^2
      \left[\frac{2}{\pi}P\!\!\int_0^\infty \!\!\!\!\! d\omega'\frac{\omega' J(\omega')}{\omega^2-\omega'^2}
    - iJ(\omega)\right]\!. \label{compJK}
\end{equation}
Comparing the imaginary parts, we have identified the spectral function of the
bath (up to prefactors) with the imaginary part of the function 
$K(\omega)$ derived in Sec.~\ref{classical} from the theory of electrical circuits,
\begin{equation}
  J(\omega) = - \mu \left(\frac{\Phi_0}{2\pi}\right)^2 {\rm Im} K(\omega).
  \label{JK}
\end{equation}
The real parts of Eq.~(\ref{compJK}) agree due to the Kramers-Kronig relation for $K(\omega)$,
\begin{equation}
  {\rm Re} K(\omega) = - \frac{2}{\pi}P\int_{0}^{\infty}\!\!\!\!\! d\omega' \frac{\omega' {\rm Im}K(\omega')}{\omega^2-\omega'^2},
\end{equation}
which can be derived from the causality relation $K(t<0)=0$, following from Eq.~(\ref{Z-FT}).

\section{Master equation}
\label{master-equation}

Starting from the quantum theory for an electrical circuit containing Josephson 
junctions and dissipative elements, Eqs.~(\ref{Hamiltonian-S}--\ref{Hamiltonian-SB}), 
we derive in this Section a generalized master 
equation for the dynamics of the Josephson phases only.
The equation of motion for the density matrix of the whole system (superconducting
phases plus reservoir modes in the external impedances) is given by the
Liouville equation,
\begin{equation}
\label{Liouville}
\dot{\rho}(t) = -i[{\cal H}, \rho(t)] \equiv  -i{\cal L}\rho(t).
\end{equation}
Following from Eq.~(\ref{Hamiltonian}), the Liouville superoperator ${\cal L}$ is the 
sum of the Liouville superoperators corresponding to the parts Eqs.~(\ref{Hamiltonian-S}), 
(\ref{Hamiltonian-B}), and (\ref{Hamiltonian-SB}) of the Hamiltonian,
${\cal L} = {\cal L}_S+{\cal L}_B+{\cal L}_{SB}$, where
${\cal L}_X \rho \equiv [{\cal H}_X, \rho]$ for $X=S,B,SB$.
In order to study the dynamics of the system without the bath, we
take the partial trace over bath modes,
\begin{equation}
  \label{rho-S}
  \rho_S(t) = {\rm Tr}_B\,\rho (t).
\end{equation}
From Eq.~(\ref{Liouville}) and with the additional assumption that
the initial state of the whole system is factorizable into a system
part $\rho_S(0)$ and an equilibrium bath part,
\begin{equation}
  \label{rho-B}
  \rho_B={\cal Z}_B^{-1}\exp(-\beta {\cal H}_B),
\end{equation}
with the bath partition function ${\cal Z}_B={\rm Tr} \exp(-\beta {\cal H}_B)$,
$\beta =1/k_B T$ being the inverse temperature,
we obtain the (exact) Nakajima-Zwanzig equation,
\begin{eqnarray}
  \dot{\rho}_S(t) &=& -i{\cal L}_S\rho_S(t) -i\int_0^t dt'\Sigma(t-t')\rho_S(t'), \label{NZ}\\
  \Sigma(t)\rho_S &=& -i {\rm Tr}_B {\cal L}_{SB} e^{-iQ{\cal L}t} {\cal L}_{SB} \rho_S\otimes\rho_B,\label{Sigma}
\end{eqnarray}
where we have used that the interaction Liouville superoperator has the form ${\cal L}_{SB}={\cal L}_{SB}^S\otimes {\cal L}_{SB}^B$ where ${\cal L}_{SB}^S$ and ${\cal L}_{SB}^B$ are system and bath superoperators, respectively, and that ${\rm Tr}_B ({\cal L}_{SB}^B \rho_B)=0$.
The projection superoperators $P$ and $Q$ are defined as
\begin{eqnarray}
  P\rho &=& \left( {\rm Tr}_B \rho\right) \otimes \rho_B\, ,\label{P}\\
  Q\rho &=& \rho - P\rho .\label{Q}
\end{eqnarray}

The Nakajima-Zwanzig equation (\ref{NZ}), with Eq.~(\ref{Sigma}),
is a formally exact and closed description
of the dynamics of the state of the system $\rho_S$, but it is rather unpractical
since it still essentially involves diagonalizing the complete problem in order
to evaluate the exponential in Eq.~(\ref{Sigma}).
However, the problem can be substantially simplified in the case of weak
coupling, i.e.\ if $||{\cal L}_{SB}|| \ll ||{\cal L}_{S}+{\cal L}_{B}||$.
We assume that the circuit contains a finite number of external impedances.
As we will see below, the weak coupling condition is satisfied here if
\begin{equation}
  \frac{J(\omega_{ij})}{\omega_{ij}}\ll 1,\quad {\rm and}\quad
  \left.\frac{J(\omega)}{\omega}\right|_{\omega\rightarrow 0} 
     \frac{k_B T}{\omega_{ij}}  \ll 1,
     \label{born-condition}
\end{equation}
hold for transition energies $\omega_{ij}$ between all possible levels $i\neq j$,
where $J(\omega)$ is given in Eq.~(\ref{JK}).
If the coupling of the external impedance is strong, $\mu\approx 1$, then 
the condition (\ref{born-condition}) requires that the involved impedance 
(resistance) is large compared to the quantum of resistance,
\begin{equation}
  \label{weak-coupling}
  Z_i, R_i \gg \frac{e^2}{h} = \frac{\pi}{2}\hbar \Phi_0^2.
\end{equation}
In the regime of Eq.~(\ref{born-condition}), 
we can expand Eq.~(\ref{Sigma}) in orders
of the system-bath interaction ${\cal L}_{SB}$.  Retaining only the
terms in first order (Born approximation) yields
\begin{equation}
\label{Born}
\Sigma_2(t)\rho_S = -i {\rm Tr}_B {\cal L}_{SB} e^{-iQ({\cal L}_S+{\cal L}_B)t} {\cal L}_{SB} \rho_S\otimes\rho_B,
\end{equation}
where the projector $Q$ in the exponent can be dropped without 
making any further approximation.

The master equation Eq.~(\ref{NZ}) in the Born approximation Eq.~(\ref{Born}), although
much simpler than the general Nakajima-Zwanzig equation, is still an integro-differential
equation that is hard to solve in general.
Further simplification is achieved with a Markov approximation
\begin{eqnarray}
\dot{\rho}_S(t) &=& -i{\cal L}_S\rho_S(t) - \Sigma_2^R(t)\rho_S(t),\label{Born-Markov}\\
\Sigma_2^R(t) &=& -i \int_0^\infty dt' \Sigma_2(t') e^{it'{\cal L}_S}.\label{Sigma-R}
\end{eqnarray}
Markov approximations rely on the assumption that the temporal correlations
in the bath are short-lived and typically lead to exponential decay of the
coherence and population.  In some situations, e.g.\ for 1/f noise,
the Markov approximation is not appropriate \cite{MSS,LeggettRMP}.
Also, note that the Markov approximation is not unique \cite{CelioLoss}.

The master equation in the Born-Markov approximation, Eqs.~(\ref{Born-Markov})
and (\ref{Sigma-R}), can be cast into the form of the Redfield equations \cite{Redfield}
by taking matrix elements in the eigenbasis $|n\rangle$ of ${\cal H}_S$
(eigenenergies $\omega_n$),
\begin{equation}
  \dot{\rho}_{nm}(t) 
   = -i\omega_{nm}\rho_{nm}(t) -\sum_{kl}R_{nmkl}\rho_{kl}(t),\label{Redfield-equation}
\end{equation}
where ${\rho}_{nm}=\langle n|\rho_S|m\rangle$, $\omega_{nm}=\omega_n-\omega_m$,
and where we have introduced the Redfield tensor,
\begin{equation}
  R_{nmkl} = \int_0^\infty \!\!\!\!\!\! dt \,{\rm Tr}_B \langle n|[{\cal H}_{SB}(t),
                  [{\cal H}_{SB}(0),|k(t)\rangle\langle l(t)|\rho_B]]|m\rangle ,
 \label{Redfield-tensor}
\end{equation}
using the interaction Hamiltonian and system eigenstates in the interaction picture,
\begin{eqnarray}
  {\cal H}_{SB}(t) &=& e^{i({\cal H}_S+{\cal H}_B)t}{\cal H}_{SB}e^{-i({\cal H}_S+{\cal H}_B)t},\\
  |k(t)\rangle &=& e^{it{\cal H}_S}|k\rangle =e^{it\omega_k}|k\rangle .
\end{eqnarray}
Further evaluation of the commutators in Eq.~(\ref{Redfield-tensor}) yields
\begin{eqnarray}
  R_{nmkl} &=& \delta_{lm}\!\sum_r \Gamma_{nrrk}^{(+)} + \delta_{nk}\!\sum_r \Gamma_{lrrm}^{(-)}
-\Gamma_{lmnk}^{(+)}-\Gamma_{lmnk}^{(-)},\nonumber\\
\label{RGamma}\\
  \Gamma_{lmnk}^{(+)} &=&  \int_0^\infty \!\!\!\!\!\!dt\, e^{-it\omega_{nk}}{\rm Tr}_B \tilde{\cal H}_{SB}(t)_{lm}\tilde{\cal H}_{SB}(0)_{nk}\rho_B,\\
\Gamma_{lmnk}^{(-)} &=&  \int_0^\infty \!\!\!\!\!\! dt\, e^{-it\omega_{lm}}{\rm Tr}_B \tilde{\cal H}_{SB}(0)_{lm}\tilde{\cal H}_{SB}(t)_{nk}\rho_B,\quad\quad\quad
\end{eqnarray}
with $\tilde{\cal H}_{SB}(t)_{nm} = \langle n| e^{it{\cal H}_B}{\cal H}_{SB}e^{-it{\cal H}_B}|m\rangle$.  Note that, using the relation $(\Gamma_{lmnk}^{(+)})^* = \Gamma_{knml}^{(-)}$,
the Redfield tensor can be expressed in terms of, e.g., the complex $\Gamma_{lmnk}^{(+)}$
tensor only.  For our system-bath interaction Hamiltonian, Eq.~(\ref{Hamiltonian-SB}),
we obtain
\begin{eqnarray}
{\rm Re}\Gamma_{lmnk}^{(+)} &=&  ({\bf m}\cdot\bphi)_{lm} ({\bf m}\cdot\bphi)_{nk} 
J(|\omega_{nk}|)\frac{e^{-\beta \omega_{nk}/2}}{\sinh \beta|\omega_{nk}|/2}\, ,\nonumber\\
{\rm Im}\Gamma_{lmnk}^{(+)} &=& -({\bf m}\cdot\bphi)_{lm} ({\bf m}\cdot\bphi)_{nk} \times \label{Gp} \\
&& \times \frac{2}{\pi} P\!\!\int_0^\infty \!\!\!\!\!\!d\omega \frac{J(\omega)}{\omega^2 \!-\!\omega_{nk}^2}\!\left(\!\omega\!-\!\omega_{nk}\coth \frac{\beta\omega}{2}\!\right). \nonumber
\end{eqnarray}

\section{Two-level approximation}
\label{two-level}

If a system is initially prepared in one of the two lowest energy 
eigenstates ($0$ and $1$) and all rates $R_{nmkl}$ for $k,l=0,1$ and 
$n,m\neq 0,1$ are negligible compared to the rates $R_{nmkl}$ for 
$n,m,k,l=0,1$ (a sufficient criterion for this being low temperature,
$\beta\omega_{12}\gg 1$), then we can restrict our description of the
system dynamics to the two lowest levels.
The 2-by-2 density matrix of the system, being Hermitian and having
trace equal to 1, can then be written in the form of three real
variables, the Bloch vector
\begin{equation}
  {\bf p} = {\rm Tr}({\mbox{\boldmath $\sigma$}}\rho ) = \left(\begin{array}{c}
      \rho_{01}+\rho_{10}\\
      i(\rho_{01}-\rho_{10})\\
      \rho_{00}-\rho_{11}
\end{array}\right),\label{Bloch-vector}
\end{equation}
where ${\mbox{\boldmath $\sigma$}}=(\sigma_x, \sigma_y, \sigma_z)$
is the vector composed of the three Pauli matrices.

By combining the Redfield equation (\ref{Redfield-equation}) with
Eq.~(\ref{Bloch-vector}), we obtain the Bloch equation,
\begin{equation}
  \label{Bloch}
  \dot{\bf p} = \bomega\times{\bf p}-R{\bf p}+{\bf p}_0,
\end{equation}
with $\bomega = (0,0,\omega_{01})^T$,
\begin{equation}
{\bf p}_0 = \left(\begin{array}{c}
    -(R_{0111}'+R_{0100}')\\
    R_{0100}''+R_{0111}''\\
    -(R_{0000}'-R_{1111}')
\end{array}\right),
\end{equation}
and the relaxation matrix
\begin{equation}
  R = \left(\!\!\begin{array}{c c c}
      R_{0101}'\!+\!R_{0110}'    &  R_{0101}'' \!-\! R_{0110}'' & R_{0100}'\!-\! R_{0111}' \\
      -R_{0101}''\!-\! R_{0110}'' &  R_{0101}' \!-\! R_{0110}'   & -R_{0100}''\!+\!R_{0111}'' \\
      2 R_{0001}'            &  R_{0001}'               & R_{0000}' \!+\! R_{1111}'
\end{array}\!\!\right),
\end{equation}
where $R_{nmkl}'={\rm Re}R_{nmkl}$ and $R_{nmkl}''={\rm Im}R_{nmkl}$.

If $\omega_{01}\gg R_{nmkl}$, we can make the secular approximation, 
only retaining terms $R_{nmkl}$ with 
$n-m = k-l$ (see e.g.\ Ref.~\onlinecite{Redfield}),
\begin{equation}
  R_{\rm sec} = \left(\begin{array}{c c c}
      R_{0101}'    &  R_{0101}''  & 0 \\
      -R_{0101}''  &  R_{0101}'   & 0 \\
      0            &  0           & R_{0000}' + R_{1111}'
\end{array}\right).
\end{equation}
The off-diagonal term $R_{0101}''$ can be absorbed into the system
Hamiltonian as a frequency renormalization,
$\tilde \omega_{01} = \omega_{01} - R_{0101}''$, and we are left with
the relaxation matrix
\begin{equation}
  \tilde{R} = \left(\begin{array}{c c c}
      T_2^{-1}    &  0           & 0 \\
      0           &  T_2^{-1}    & 0 \\
      0           &  0           & T_1^{-1}
\end{array}\right),
\end{equation}
where the relaxation and decoherence times are given by
\begin{eqnarray}
  \frac{1}{T_1}  &=& R_{0000}' + R_{1111}' = 2{\rm Re}(\Gamma^{(+)}_{0110}+\Gamma^{(+)}_{1001}),\\
  \frac{1}{T_2}  &=& R_{0101}'             = \frac{1}{2 T_1} 
 + {\rm Re}(\Gamma^{(+)}_{0000}+\Gamma^{(+)}_{1111}-2 \Gamma^{(+)}_{0011}) \nonumber\\ 
 &=& \frac{1}{2 T_1} + \frac{1}{T_\phi},\\
  \frac{1}{T_\phi}  &=& {\rm Re}(\Gamma^{(+)}_{0000}+\Gamma^{(+)}_{1111}-2 \Gamma^{(+)}_{0011}).
\end{eqnarray}
Using Eq.~(\ref{Gp}), we obtain
\begin{eqnarray}
  \frac{1}{T_1} &=& 4|\langle 0|{\bf m}\cdot\bphi|1\rangle|^2 J(\omega_{01}) \coth\frac{\omega_{01}}{2k_B T}, \label{T1}\\
  \frac{1}{T_\phi} &=&  |\langle 0|{\bf m}\cdot\bphi|0\rangle-\langle 1|{\bf m}\cdot\bphi|1\rangle|^2 \left.\frac{J(\omega)}{\omega}\right|_{\omega\rightarrow 0} \!\!\!\!\!\!\!\!\! 2k_B T. \quad\quad\label{Tphi}
\end{eqnarray}
Typically, $T_\phi$ can be made to diverge by changing the external fluxes
until $\langle 0|{\bf m}\cdot\bphi|0\rangle = \langle 1|{\bf m}\cdot\bphi|1\rangle$.
It can be expected, however, that this divergence will be cut off by effects that
are beyond the present theory, e.g.\ other noise sources, higher-order corrections,
or non-Markovian effects.

\subsection{Semiclassical approximation}
\label{semiclassical}

Let us assume that the potential $U(\bphi)$ describes a double well
with ``left'' and ``right'' minima at $\bphi_L$ and $\bphi_R$.  
Furthermore, for the moment we make a semiclassical approximation in which
the left and right single-well groundstates $|L\rangle$ and $|R\rangle$ 
centered  at $\bphi_{L,R}$ are localized orbitals, i.e.\ they do not overlap each other.
Then the two lowest eigenstates can approximately be written as
the symmetric and antisymmetric combinations of $|R\rangle$ and $|L\rangle$,
\begin{eqnarray}
  |0\rangle  &=&  \frac{1}{\sqrt{2}}\left(\sqrt{1+\frac{\epsilon}{\omega_{01}}}\,|L\rangle + \sqrt{1-\frac{\epsilon}{\omega_{01}}}\,|R\rangle\right),\label{TL0}\\
  |1\rangle  &=&  \frac{1}{\sqrt{2}}\left(\sqrt{1-\frac{\epsilon}{\omega_{01}}}\,|L\rangle - \sqrt{1+\frac{\epsilon}{\omega_{01}}}\,|R\rangle\right),\label{TL1}
\end{eqnarray}
where $\omega_{01}=\sqrt{\Delta^2 +\epsilon^2}$, 
$\Delta = \langle L|{\cal H}_S|R\rangle$ is the tunneling amplitude between the two wells,
and $\epsilon = \langle L|{\cal H}_S|L\rangle - \langle R|{\cal H}_S|R\rangle$ 
the asymmetry of the double well.
Since $|L\rangle$ and $|R\rangle$ are localized orbitals,
we can approximate
\begin{equation}
  \langle L|\bphi|R \rangle \approx  0,\quad
  \langle L|\bphi|L \rangle \approx  \bphi_L,\quad
  \langle R|\bphi|R \rangle \approx  \bphi_R.\label{SC-matrix}
\end{equation}
From Eqs.~(\ref{TL0})--(\ref{SC-matrix}) the eigenstate matrix elements are
\begin{eqnarray}
  \langle 0|\bphi|1\rangle &\approx & \frac{1}{2}\frac{\Delta}{\omega_{01}} \Delta\bphi,\\
  \langle 0|\bphi|0\rangle-\langle 1|\bphi|1\rangle &\approx & \frac{\epsilon}{\omega_{01}} \Delta\bphi,
\end{eqnarray}
where $\Delta\bphi = \bphi_L-\bphi_R$.
Finally, the relaxation and pure dephasing times for a
double-well potential in the semiclassical limit becomes
\begin{eqnarray}
  \frac{1}{T_1} &=& \left(\frac{\Delta}{\omega_{01}}\right)^2\left| \Delta\bphi \cdot{\bf m}\right|^{2} J(\omega_{01}) \coth\frac{\omega_{01}}{2k_B T}, \label{T1-dw}\\
  \frac{1}{T_\phi} &=&  \left(\frac{\epsilon}{\omega_{01}}\right)^2 |\Delta\bphi \cdot {\bf m}|^2 \left.\frac{J(\omega)}{\omega}\right|_{\omega\rightarrow 0} \!\!\!\!\!\!\!\!\! 2k_B T. \quad\quad\label{Tphi-dw}
\end{eqnarray}
In this semiclassical approximation with localized states,
the relaxation and decoherence times both diverge if $\Delta\bphi$ can
be made orthogonal to ${\bf m}$.  For a symmetric double well ($\epsilon =0$),
$T_\phi\rightarrow\infty$ for all $\Delta \bphi$.

\subsection{Quantum corrections}
\label{qcorr}

Quantum corrections to the semiclassical approximation discussed 
in Sec.~\ref{semiclassical} can be estimated by taking into account the
finite spread of the wavefunction about its center, using a
(approximate) quadratic Hamiltonian at the potential minimum
\begin{equation}
  H = \frac{1}{2} \left({\bf Q}_C ^T {\bf C}^{-1} {\bf Q}_C 
       + \left(\frac{\Phi_0}{2 \pi}\right)^2 \bphi ^T {\bf L}_{\rm lin}^{-1} \bphi \right),
\end{equation}
where
\begin{equation}
  {\bf L}_{\rm lin}^{-1}={\bf M}_0+{\rm diag}\left(\frac{\cos(\varphi_{L,R;i})}{L_{J;i}}\right),
\end{equation}
Rescaling $\bphi$ and its conjugate momentum ${\bf Q}_C$,
\begin{eqnarray}
  \bphi &=& \frac{2 \pi}{\Phi_0}\sqrt{{\bf C}}^{-1}\tilde\bphi,\quad
  \left[\varphi_i = \frac{1}{\sqrt{C_i}}\frac{2 \pi}{\Phi_0}\tilde\varphi_i\right] , \\
  {\bf Q}_C  &=& \sqrt{{\bf C}} \tilde{\bf Q}_C, \quad
  \left[Q_{C;i} = \sqrt{C_i}\tilde Q_{C;i}\right] ,
\end{eqnarray}
we obtain the Hamiltonian
\begin{equation}
  H = \frac{1}{2}\left(\tilde{\bf Q}_C ^2 + \tilde\bphi ^T {\bf \Omega} \tilde\bphi \right)
    = \frac{1}{2}\left(\tilde{\bf Q}_C ^2 +\sum_i\omega_i^2 ({\mbox{\boldmath $\xi$}}_i\cdot\tilde\bphi)^2\right),
\end{equation}
where the inverse LC-matrix is defined as
\begin{equation}
  {\bf \Omega}  = \left(\sqrt{{\bf C}}^{-1}\right)^T {\bf L}_{\rm lin}^{-1} \sqrt{{\bf C}}^{-1},
\label{qH}
\end{equation}
and we have diagonalized the ${\bf \Omega}$ matrix,
${\bf \Omega}{\mbox{\boldmath $\xi$}}_i=\omega_i^2{\mbox{\boldmath $\xi$}}_i$.
The ground-state wavefunction of the harmonic oscillator Hamiltonian Eq.~(\ref{qH})
is a Gaussian centered at the left (L) or right (R) potential minimum,
\begin{widetext}
\begin{equation}
  \Psi(\bphi) = \langle \bphi |L,R\rangle = \left(\frac{\Phi_0}{2\pi}\right)^{N_J/2}
                 \prod_{i=1}^{N_J} \left(\frac{C_i\omega_i}{\pi \hbar}\right)^{1/4}
                 \exp\left( - \sum_i \omega_i \left(\frac{\Phi_0}{2\pi}\right)^2
                               \left({\mbox{\boldmath $\xi$}}_i^T{\bf C}(\bphi-\bphi_{L,R})\right)^2\right).
\end{equation}
\end{widetext}
The wavefunction overlap integral between the left and right state is found to be
\begin{equation}
S \equiv \langle L| R\rangle = \exp\left(-\frac{1}{2\hbar}\left(\frac{\Phi_0}{2\pi}\right)^2 \sum_{i=1}^{N_J} \omega_i ({\mbox{\boldmath $\xi$}}_i^T{\bf C}\Delta\bphi)^2\right),\label{overlap}
\end{equation}
Note that in the classical limit, where all capacitances $C_i$ are large, the
overlap tends to zero, $\langle L|R\rangle\rightarrow 0$. 
Introducing the orthogonalized (Wannier) orbitals,
\begin{eqnarray}
  |\tilde L\rangle &=& \frac{|L\rangle -g|R\rangle}{\sqrt{1-2Sg+g^2}},\\
  |\tilde R\rangle &=& \frac{|R\rangle -g|L\rangle}{\sqrt{1-2Sg+g^2}},\\
  g &=& \frac{1-\sqrt{1-S}}{S},
\end{eqnarray}
we can derive the matrix elements,
\begin{eqnarray}
  \langle \tilde L|\bphi |\tilde R\rangle &=& 0,\\
  \langle \tilde L|\bphi |\tilde L\rangle &=& \frac{(1-g^2)\bphi _L +2g(g-S)\bphi _0}{1-2Sg+g^2},\\
  \langle \tilde R|\bphi |\tilde R\rangle &=& \frac{(1-g^2)\bphi _R +2g(g-S)\bphi _0}{1-2Sg+g^2},
\end{eqnarray}
and the difference,
\begin{equation}
  \langle \tilde L|\bphi |\tilde L\rangle - \langle \tilde R|\bphi |\tilde R\rangle 
   =  \frac{1-g^2}{1-2Sg+g^2}\Delta\bphi
   \approx  \left(1+\frac{S^2}{2}\right)\Delta\bphi ,
\end{equation}
where $S$ is defined in Eq.~(\ref{overlap}).
By replacing $|L\rangle$ and $|R\rangle$ by $|\tilde L\rangle$ and $|\tilde R\rangle$ 
in Eqs.~(\ref{TL0}) and (\ref{TL1}), we obtain
\begin{eqnarray}
  \langle 0|\bphi|1\rangle &\approx & \frac{1}{2}\frac{\Delta}{\omega} \left(1+\frac{S^2}{2}\right)\Delta\bphi,\\
  \langle 0|\bphi|0\rangle-\langle 1|\bphi|1\rangle &\approx & \frac{\epsilon}{\omega} \left(1+\frac{S^2}{2}\right)\Delta\bphi .
\end{eqnarray}
Note that in this semiclassical approximation using Gaussian orbitals, both
$T_1$ and $T_\phi$, Eqs.~(\ref{T1-dw}) and (\ref{Tphi-dw}), and thus also $T_2$, 
are renormalized by a factor $(1+S^2/2)^{-1}$, but for
the symmetric double-well ($\epsilon=0$), $T_\phi$
is still infinite.

\section{Leakage}
\label{leakage}

We can go beyond the two-level approximation, e.g., by looking at the 
leakage out of the two lowest levels.  
Within the secular approximation, 
the total rates for transition out of the allowed qubit states $|k\rangle$ ($k=0,1$) can
be written as
\begin{equation}
  \frac{1}{T_{L,k}} = 4\sum_n |\langle n|{\bf m}\cdot\bphi|k\rangle|^2 J(\omega_{kn}) \coth\frac{\omega_{kn}}{2k_B T}. \label{Tleak}
\end{equation}

As an example, we model leakage by
adding two additional levels $|2\rangle$ and $|3\rangle$ to the allowed 
logical qubit states $|0\rangle$ and $|1\rangle$ and derive the typical rate for
transitions from $|0,1\rangle$ to $|2,3\rangle$ due to the coupling to the
environment.
In analogy to Eqs.~(\ref{TL0}) and (\ref{TL1}), 
the excited states originating from two coupled 
single-well excited states $|L'\rangle$ and $|R'\rangle$ can be written as
\begin{eqnarray}
  |2\rangle  &=&  \frac{1}{\sqrt{2}}\left(\sqrt{1+\frac{\epsilon}{\omega_{23}}}\,|L'\rangle + \sqrt{1-\frac{\epsilon}{\omega_{23}}}\,|R'\rangle\right),\quad\quad\label{TL2}\\
  |3\rangle  &=&  \frac{1}{\sqrt{2}}\left(\sqrt{1-\frac{\epsilon}{\omega_{23}}}\,|L'\rangle - \sqrt{1+\frac{\epsilon}{\omega_{23}}}\,|R'\rangle\right),\quad\quad\label{TL3}
\end{eqnarray}
where $\omega_{23}=\sqrt{\Delta'^2 +\epsilon^2}$ 
and $\langle L|L'\rangle = \langle R|R'\rangle = 0$.  
We model the coupling to the lowest two levels by the perturbation Hamiltonian
\begin{equation}
  H' = -\delta \left( |L\rangle\langle R'| + |R\rangle\langle L'|  + h.c.\right),
\end{equation}
and denote the energy splitting between the
lowest two states $|L\rangle$, $|R\rangle$ and the higher energy states 
$|L'\rangle$ and $|R'\rangle$ with $\eta$.
In the regime $\eta \gg \Delta, \delta, \epsilon \gg \Delta'$,
the matrix elements of the phase coordinate $\bphi$
in the coupled states $|\tilde n\rangle$ are found to be
\begin{eqnarray}
  \langle \tilde 0|\bphi|\tilde 3\rangle &\approx & -\langle \tilde 1|\bphi|\tilde 2\rangle 
    \approx  \frac{1}{\sqrt{2}}\frac{\delta}{\eta} \sqrt{1+\frac{\epsilon}{\omega}}\Delta\bphi,\\
  \langle \tilde 0|\bphi|\tilde 2\rangle &\approx &  \langle \tilde 1|\bphi|\tilde 3\rangle 
    \approx    \frac{1}{\sqrt{2}}\frac{\delta}{\eta} \sqrt{1-\frac{\epsilon}{\omega}}\Delta\bphi .
\end{eqnarray}
The dominant leakage occurs with the rate
\begin{equation}
  \frac{1}{T_L} \approx  4\left(\frac{\delta}{\eta}\right)^2 |{\bf m}\cdot\Delta\bphi|^2
  J(\eta) \coth\frac{\eta}{2k_B T}.
  \label{leakage-TL}
\end{equation}
Note that (thermally activated) leakage is not relevant if $T\ll \eta$,
in spite of a finite rate $T_L^{-1}$, because the population of the
excited states in thermal equilibrium is exponentially suppressed.

\section{The IBM qubit}
\label{ibm-qubit}

In this Section, we use the theory developed in Secs.~\ref{classical}--\ref{two-level}
to describe decoherence and relaxation in a superconducting flux qubit design which is 
currently under experimental study by a group at IBM \cite{IBM}.
This superconducting circuit resembles a dc SQUID, with one Josephson 
junction replaced by another dc SQUID, see Fig.~\ref{ibm-graph}.  
The circuit thus comprises three Josephson junctions in total.  
This design has the advantage that it provides a high level of control.
There are three externally adjustable parameters;  the external magnetic fluxes 
threading the larger (main) loop and the smaller (control) loop, and the bias
current $I_B$.

\subsection{Current biased circuit}
\label{ibm-qubit-current}

We first study the decoherence due to a current source that is attached
to the circuit, see Fig.~\ref{ibm-graph}.  It is unavoidable that
the external current source will also introduce a coupling to 
an external impedance $Z$.  In our model, this impedance is connected
in parallel with an ideal current source.  The impedance $Z(\omega)$
as a function of frequency $\omega$ can be determined experimentally \cite{IBM}.

\begin{figure}
\centerline{\includegraphics[width=8cm]{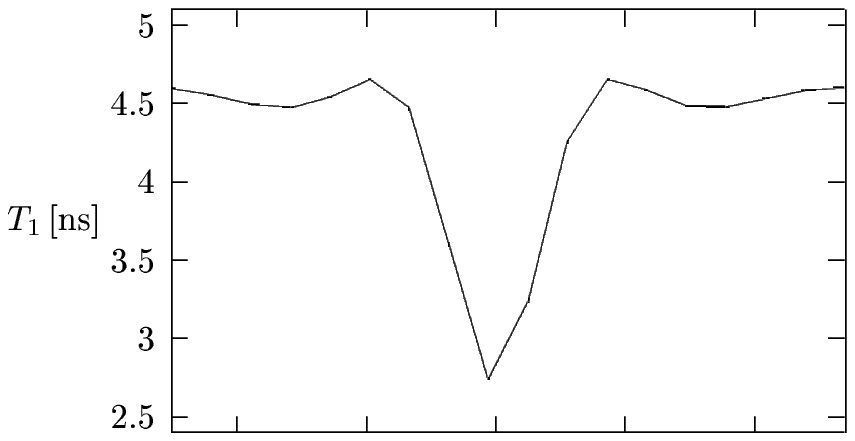}}
\centerline{\includegraphics[width=8cm]{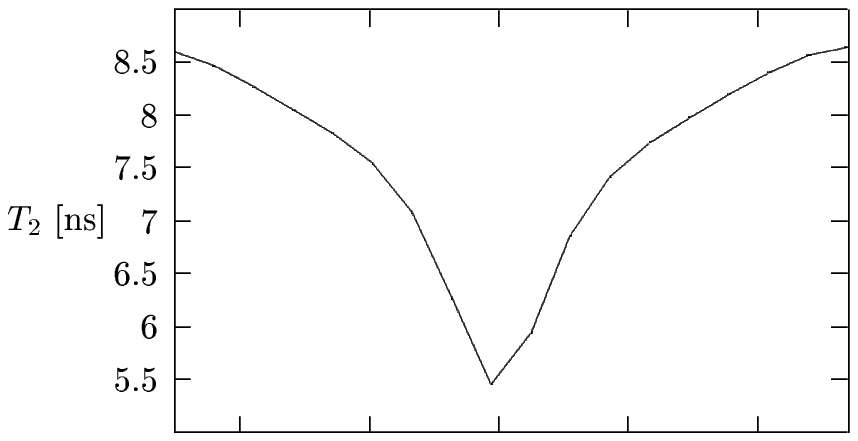}}
\centerline{\includegraphics[width=8cm]{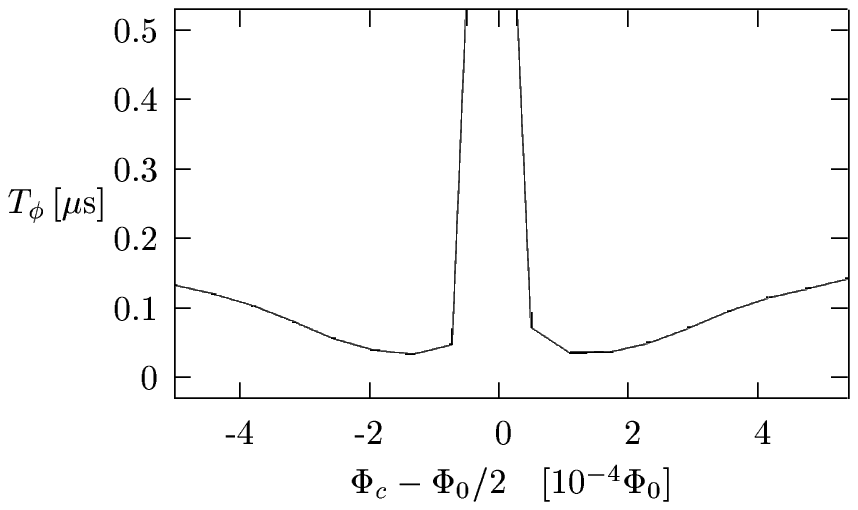}}
\caption{\label{ibm-T}
Relaxation time $T_1$, decoherence time $T_2$, and 
pure dephasing time $T_\phi$ for the 
current-biased IBM qubit as a function of the control flux $\Phi_c$
around the point $\Phi_c=\Phi_0/2$.
The main flux $\Phi$ is chosen such that the resulting 
double-well is always symmetric.
While $T_\phi$ diverges at the point $\Phi_c = \Phi_0/2$ where 
$\langle 0|{\bf m}\cdot\bphi|0\rangle = \langle 1|{\bf m}\cdot\bphi|1\rangle$,
$T_1$ has a minimum at that point.
The inductances for this example are 
$L_1=L_4=100\,{\rm pH}$ (main loop) and  $L_2=L_3=4\,{\rm pH}$ (control loop).
The capacitance and critical current of the junctions are $C=0.1\,{\rm pF}$ and
$I_c=8.5\,\mu{\rm A}$ ($L_J=\Phi_0/2\pi I_c=39\,{\rm pH}$).
The external impedance is assumed to be $Z(0)=2.5\,{\rm k}\Omega$ at zero
frequency and $Z(\omega_{01})=10\,{\rm k}\Omega$ at
the transition frequency $\omega_{01}$;  the temperature
of the external impedance is taken to be $30\,{\rm mK}$.}
\end{figure}

We choose the tree shown in Fig.~\ref{ibm-tree} for the graph 
representing the IBM circuit ($N=6$ nodes and $B=15$ branches) and 
obtain the following network graph characteristics (cf.\ Sec.~\ref{classical}),
\begin{equation}
  \label{FC-ibm-c}
  {\bf F}_{CL} = \left(\begin{array}{r r}
      1 &  0 \\
     -1 &  1 \\
      0 & -1
\end{array}\right), \quad 
  {\bf F}_{CZ} = -{\bf F}_{CB} = \left(\begin{array}{r}
      0 \\
      1 \\
      0
\end{array}\right),
\end{equation}
\begin{equation}
  \label{FK-ibm-c}
  {\bf F}_{KL} = \left(\begin{array}{r r}
     -1 &  1 \\
     -1 &  0
\end{array}\right),\quad
  {\bf F}_{KZ} = -{\bf F}_{KB} = \left(\begin{array}{r}
      1 \\
      1
\end{array}\right).
\end{equation}
The linear inductances are given by
\begin{equation}
  \label{inductance-ibm-c}
  {\bf L} =   \left(\begin{array}{c c}
     L_1 & 0 \\
     0   & L_3
\end{array}\right)\! , \:\: 
  {\bf L}_K =   \left(\begin{array}{c c}
     L_2 & 0 \\
     0   & L_4
\end{array}\right)\! , \:\: 
  {\bf L}_{KL} = {\bf 0}.
\end{equation}
Using Eqs.~(\ref{M0-s}) and (\ref{N-s})
with Eqs.~(\ref{FC-ibm-c}), (\ref{FK-ibm-c}), (\ref{inductance-ibm-c}),
we obtain the parameters for the Hamiltonian,
\begin{eqnarray}
  {\bf M}_0 &=&
  \frac{1}{B_0} \left(\!\begin{array}{c c c}
      L_2 \!+\! L_3 & -L_3           & -L_2 \\
     -L_3      &  L_1 \!+\! L_3 \!+\! L_4 & -L_1 \!-\! L_4 \\
     -L_2      & -L_1 \!-\! L_4      & L_1 \!+\! L_2 \!+\! L_4
\end{array}\!\right)\!,\label{M0-ibm}\\ 
  {\bf N} &=&
  \frac{1}{B_0} \!\left(\begin{array}{c c}
     L_2+L_3   & L_2        \\
     -L_3      & L_1+L_4    \\
     -L_2      & -L_1-L_2-L_4
\end{array}\right)\!,\hspace{2.2cm} \label{N-ibm}
\end{eqnarray}
where
\begin{equation}
B_0=L_1 L_2  + L_1 L_3  + L_2 L_3  + L_2 L_4  + L_3 L_4.\label{B0}
\end{equation}
For the dissipative part, we use Eq.~(\ref{mbar-s}) and Eqs.~(\ref{K-def})--(\ref{m}), 
with the result
\begin{equation}
  \label{K-ibm-current}
  K(\omega)= B_0 / B_\omega ,
\end{equation}
where $B_\omega = L_1 L_2 L_3 + L_1 L_2 L_4 + L_1 L_3 L_4 + L_1 L_2 L_z(\omega) + L_1 L_3 L_z(\omega) + L_2 L_3 L_z(\omega) + L_2 L_4 L_z(\omega) + L_3 L_4 L_z(\omega)$,
which allows us to determine the spectral density $J(\omega)$ of the
bath using Eq.~(\ref{JK}), and 
\begin{eqnarray}
  \mu  &=& \frac{L_1^2 (L_2^2 + L_3^2) + (L_3 L_4 + L_2 (L_3 + L_4))^2}{B_0^2},\label{IBM-c-mu}\\
  {\bf m} &=&  \frac{1}{B_0 \sqrt{\mu}}
  \left(\begin{array}{c} L_3 L_4+L_2 (L_3 +L_4) \\  L_1 L_3 \\  L_1 L_2
        \end{array}\right) .\label{IBM-c-m}
\end{eqnarray}
Since the bias current in shunted in parallel to the external impedance,
we find ${\bf S} = - \bar{\bf m}= - \sqrt{\mu} {\bf m}$.
We can further simplify the expressions in the case of
symmetric loops, $L_4=L_1$ and $L_3=L_2$,
\begin{eqnarray}
  K(\omega) &=& \frac{4 L_1 + L_2}{2L_1^2 +L_1 L_2 +4 L_1 L_z +L_2 L_z},\label{IBM-c-K-sym}\\
  \mu       &=& \frac{6 L_1^2 +4 L_1 L_2 +L_2^2}{(4 L_1 + L_2)^2},\label{IBM-c-mu-sym}\\
  {\bf m}   &=& \frac{1}{\sqrt{6 L_1^2 + 4 L_1 L_2 + L_2^2}}
  \left(\begin{array}{c} 2 L_1 +L_2  \\  L_1 \\  L_1
        \end{array}\right) .\quad \label{IBM-c-m-sym}
\end{eqnarray}
Moreover, if the control loop inductance is much smaller than 
the main loop inductance, $L_1\gg L_2$, we obtain the asymptotics
\begin{eqnarray}
  K(\omega) &\approx & \frac{1}{Z(\omega)/i\omega + L_1/2} \approx \frac{i\omega}{Z(\omega)},
\label{IBM-c-K-s}\\
  \mu  &\approx & \frac{3}{8},\label{IBM-c-mu-s}\\
  {\bf m}  &\approx & \frac{1}{\sqrt{6}}
  \left(\begin{array}{c} 2 \\  1 \\  1
        \end{array}\right).\label{IBM-c-m-s}
\end{eqnarray}
The second approximation for $K(\omega)$ is suitable if $\omega\ll Z(\omega)/L_1$
which holds for $\omega \ll \omega_{LR} \approx 150\,{\rm GHz}$ for $Z\approx 100\,\Omega$ and $L_1\approx 100\,{\rm pH}$.
In Fig.~\ref{ibm-T}, the relaxation and decoherence times $T_1$ and $T_2$ in this regime 
are plotted as a function of the externally
applied magnetic flux $\Phi_c$, using a numerical solution of ${\cal H}_S$.

There is an intuitive explanation for the simplified result 
Eq.~(\ref{IBM-c-m-s});
both an external bias current $I_B$ and the current 
fluctuations from the external impedance $Z$ are split equally between the 
right and left half of the main loop (the two halves having equal inductances).
For this splitting of the current (fluctuations), the inductance of the
control loop is irrelevant, since it is negligible compared to the inductance of the
main loop.
The current in the left half of the main loop is further split equally between
the two halves of the control loop (having equal inductances).
Thus, the ratio of current (fluctuations) flowing through each of the
Josephson junctions is 2:1:1, which is reflected in the coupling vector
${\bf m}$ for current fluctuations from the bath to the superconducting phases
$\bphi$ pertaining to the Josephson junctions in the right half of the main loop,
and the right and left halves of the control loop, and also in the vector ${\bf S}$
describing the coupling of an external current to the superconducting phases.

\subsection{Flux biased circuit}

\begin{figure}
\centerline{\includegraphics[width=7.55cm]{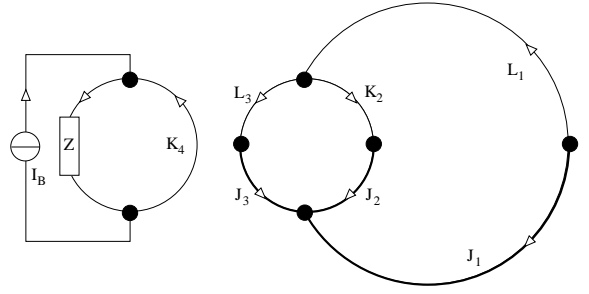}}
\caption{\label{ibm-flux}
The flux-biased IBM qubit.  The coil inductance $K_4$ can either be
coupled to the main loop via a mutual inductance to $L_1$ or to the
control loop via $L_3$.}
\end{figure}

Further control for the system shown in Fig.~\ref{ibm-graph} in addition to a current bias 
line can is achieved by inductively changing the magnetic flux through the two loops,
see Fig.~\ref{ibm-flux}.  This type of control also potentially introduces decoherence due
to fluctuations of the external fluxes.  Another way of looking at this effect would be
to say that, again, current fluctuations are caused by an external impedance in the coil
producing the flux; subsequently, these current fluctuations are transferred to the superconducting
circuit via a ``transformer'', i.e.\ via the mutual inductance between the coil and the 
superconducting qubit.  As in the case of the external bias current,  the decoherence 
processes are unavoidable if external control is to be applied.
The method introduced above can be used in the same way as before to derive the Hamiltonian
and the spectral density and form of coupling of the dissipative environment.
The network graph ($N=7$ nodes, $B=15$ branches) shown in Fig.~\ref{ibm-flux}
has the following characteristics,
\begin{eqnarray}
  {\bf F}_{CL} = \left(\begin{array}{r r}
      1 &  0 \\
     -1 &  1 \\
      0 & -1
\end{array}\right), & &
  {\bf F}_{CZ} = {\bf F}_{CB} = {\bf 0}. \label{FC-ibm-f} \\
  {\bf F}_{KL} = \left(\begin{array}{r r}
     -1 &  1 \\
      0 &  0
\end{array}\right), & &
  {\bf F}_{KZ} = -{\bf F}_{KB} = \left(\begin{array}{r}
      0 \\
     -1
\end{array}\right). \quad\quad\label{FK-ibm-f}
\end{eqnarray}
The structure of the inductance matrix depends on whether
the external flux is coupled to the main loop or the control loop.

\subsubsection{Main flux bias}
For an external coil coupled to the main (larger) loop,
the inductances are
\begin{eqnarray}
  {\bf L} =   \left(\begin{array}{c c}
     2 L_1 & 0 \\
     0   & L_3
\end{array}\right), & \quad\quad &
  {\bf L}_K =   \left(\begin{array}{c c}
     L_2 & 0 \\
     0   & L_c
\end{array}\right), \nonumber\\
  {\bf L}_{LK} =   \left(\begin{array}{c c}
     0   & M \\
     0   & 0
\end{array}\right),& & \label{inductance-ibm-f-m}
\end{eqnarray}
where $L_c$ denotes the self-inductance of the coil and
$M$ the mutual inductance between the coil and the main loop.

Since the system without external coupling is the same as for the current-biased version,
the system Hamiltonian, i.e.\ the expressions for ${\bf M}_0$, ${\bf N}$, 
and $B_0$ are the same as for the current-biased circuit, Eqs.~(\ref{M0-ibm}) and
(\ref{N-ibm}), with $L_4=L_1$.
The spectral density is obtained via Eq.~(\ref{JK}) and the result
$K(\omega)=B_0/B_\omega$
where $B_\omega=2 L_1 (L_2 + L_3)(L_c + L_z)- L_3 M^2 + L_2(L_3 (L_c + L_z)) - M^2)$.
For $\mu$ and ${\bf m}$ we find
\begin{eqnarray}
  \mu  &=& M^2\frac{2 (L_2^2 + L_2 L_3 + L_3^2)}{(L_2 L_3 + 2 L_1 (L_2 + L_3))^2},\\
  {\bf m}  &=& \frac{1}{\sqrt{L_2^2 + L_3^2 +(L_2+L_3)^2}}
  \left(\begin{array}{c} -(L_2+L_3)  \\  L_3 \\  L_2
        \end{array}\right) .\quad\quad
\end{eqnarray}
We study the following special cases of this result.
If the control loop is symmetric, $L_3=L_2$, we obtain the simpler expressions
\begin{eqnarray}
  K(\omega) &=& \frac{4 L_1 + L_2}{4 L_1 (L_c+L_z)+L_2(L_c+L_z)-2 M^2},\quad\\
  \mu       &=& \frac{6 M^2}{(4 L_1 + L_2)^2},\quad\quad
  {\bf m}   = \frac{1}{\sqrt{6}}
  \left(\begin{array}{c} -2  \\  1 \\  1 \label{IBM-fm-m-s}
        \end{array}\right) .
\end{eqnarray}
If for a symmetric control loop, 
the control loop inductance is much smaller than the main loop inductance, i.e.\ 
for $L_1\gg L_2$, we find
\begin{eqnarray}
  K(\omega) &=& \frac{4 L_1}{4 L_1 (L_c+L_z)-2 M^2} \approx \frac{i\omega}{Z(\omega)},\\
  \mu       &=& \frac{3 M^2}{8 L_1^2} .
\end{eqnarray}
The second approximation for $K(\omega)$ is suitable if $\omega\ll Z(\omega)/L_c , Z(\omega) L_1/M^2$.
The intuitive explanation for the result Eq.~(\ref{IBM-fm-m-s}) is essentially the same
as above for Eq.~(\ref{IBM-c-m-s}), with the difference that the inductively coupled 
current fluctuations couple oppositely to the Josephson junction in the main loop.

\subsubsection{Control flux bias}

An external coil coupled to the control (small) loop can be
described by the inductances
\begin{eqnarray}
  \label{inductance-ibm-f-c}
  {\bf L} =   \left(\!\begin{array}{c c}
     2 L_1 & 0 \\
     0   & L_3
\end{array}\!\right), & &
  {\bf L}_K =   \left(\!\begin{array}{c c}
     L_2 & -M/2 \\
    -M/2 & L_c
\end{array}\!\right), \quad\quad\\
  {\bf L}_{LK} =   \left(\!\begin{array}{c c}
     0   & 0 \\
     0   & M/2
\end{array}\!\right), & &
\end{eqnarray}
where $L_c$ is the self-inductance of the coil and $M$ is the total 
mutual inductance between the coil and the control loop.
Again, the expressions for ${\bf M}_0$, ${\bf N}$, and $B_0$ are
the same as for the current-biased circuit.
We find $K(\omega)=B_0/B_\omega$ 
with $B_\omega=(-L_3 M^2+8 L_1(L_2(L_c+L_z)+L_3(L_c+L_z)-M^2)+L_2(4 L_3(L_c+L_z)-M^2))/4$,
and
\begin{eqnarray}
  \mu      &=& M^2\frac{16 L_1^2+L_2^2-L_2 L_3+L_3^2+4 L_1(L_2+L_3)}{2(L_2 L_3+2 L_1(L_2+L_3))^2},\nonumber\\
\\
  {\bf m}  &=& \frac{1}{\sqrt{(4 L_1+L_2)^2 + (4 L_1 +L_3)^2 + (L_2-L_3)^2}}\times\nonumber\\
               && \times\left(\begin{array}{c} -L_2+L_3  \\  -(4 L_1+L_3) \\  4 L_1 +L_2
        \end{array}\right).
\end{eqnarray}
Since the bias current is shunted parallel to the external impedance, we find
${\bf S} = -\bar{\bf m}=-\sqrt{\mu}{\bf m}$.
For a symmetric control loop, $L_3=L_2$, we obtain
\begin{eqnarray}
  K(\omega) &=& \frac{2 L_2}{2 L_2 (L_c+L_z)- M^2} \approx \frac{i\omega}{Z(\omega)},\\
  \mu       &=& \frac{M^2}{2 L_2^2},\quad\quad
  {\bf m}   = \frac{1}{\sqrt{2}} \left(\begin{array}{c} 0  \\ -1  \\ 1  
        \end{array}\right) .\label{IBM-fc-m-s}
\end{eqnarray}
The second approximation for $K(\omega)$ is suitable if 
$\omega\ll Z(\omega)/L_c , Z(\omega) L_2/M^2$.

The result Eq.~(\ref{IBM-fc-m-s}) reflects the fact that
in the symmetric case, $L_3=L_2$,
a control flux bias only affects the superconducting phases
in the control loop.  The two phases are affected with the
same magnitude of fluctuations, but with opposite sign.

\section{The Delft qubit}
\label{delft-qubit}

As a further application of our theory, we study decoherence in
a superconducting circuit studied experimentally as a candidate
for a superconducting flux qubit in Refs.~\onlinecite{Mooij,Orlando}.
\begin{figure}[b]
\centerline{\includegraphics[width=8cm]{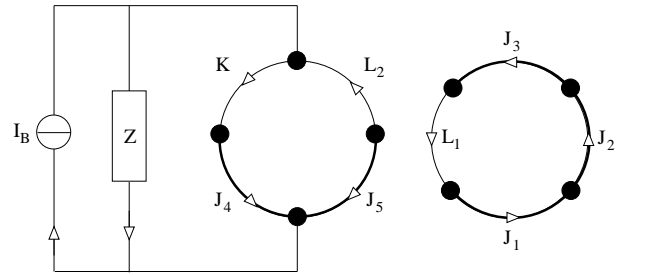}}
\caption{\label{delft-graph}
The graph representation of the Delft qubit. The ``qubit'' loop (right) involves three
Josephson junctions $J_i$ (i=1,2,3; thick lines) 
and is inductively coupled to the SQUID (or read-out) 
loop (left) which comprises two junctions and can be current biased.}
\end{figure}
The circuit consists of a ring similar to a dc SQUID but with three
junctions, see Fig.~\ref{delft-graph}.  For readout, a dc SQUID is inductively
coupled to the three-junction (qubit) loop.  The readout SQUID is current biased in
order to find its critical current.  The value of the critical current can
then be used to determine the state of the qubit loop.

This circuit network graph characteristics of the Delft qubit are ($N=8$ nodes, $B=20$
branches),
\begin{eqnarray}
  {\bf F}_{CL} = \left(\begin{array}{r r}
      -1 &  0 \\
      -1 &  0 \\
      -1 &  0 \\
       0 & -1 \\
       0 &  1
\end{array}\right), & &
  {\bf F}_{CZ} = -{\bf F}_{CB} = \left(\begin{array}{r}
      0 \\
      0 \\
      0 \\
      1 \\
      0
\end{array}\right)\!,  \label{FC-delft}\\
  {\bf F}_{KL} = \left(\begin{array}{r r}
     0 & -1 
\end{array}\right), & &
  {\bf F}_{KZ} = -{\bf F}_{KB} = \left(\begin{array}{r}
      1
\end{array}\right)\!. \quad\quad\label{FK-delft}
\end{eqnarray}
We use the following assignment for the inductances
\begin{equation}
  {\bf L} =   \left(\!\begin{array}{c c}
     L   & M_R \\
     M_R & L_R
\end{array}\!\right)\!,\,
  {\bf L}_K =   \left(\!\begin{array}{c c}
     L_L
\end{array}\!\right)\!,\,
  {\bf L}_{LK} =   \left(\!\begin{array}{c}
     M_L\\
     M'
\end{array}\!\right)\!,\, \label{inductance-delft}
\end{equation}
where $L$ and $L'=L_L+L_R$ are the self-inductances of the qubit and SQUID
loops, respectively, and $M=M_L+M_R$ is the mutual inductance between the
two loops.  The self-inductance of the SQUID loop and the mutual inductance 
are divided into parts $L_L$ and $M_L$ corresponding to the left half of the
SQUID loop and parts $L_R$ and $M_R$ corresponding to the right half of the SQUID
loop.
We introduce the following notations and conventions.
The Josephson inductances of the five junctions are given by
$L_{J,1}=L_{J,2}=L_J$ and $L_{J,3}=L_J/\beta$ for the three qubit junctions, 
and $L_{J;L,R}=L_J'$ for the two SQUID junctions.
The superconducting phase differences across the five junctions are denoted
with $\bphi=(\varphi_1, \varphi_2, \varphi_3, \varphi_L, \varphi_R)$,
and the capacitances of the five junctions are 
${\bf C}={\rm diag}(C,C,C,C',C')$.
The externally applied fluxes threading the qubit and SQUID loops are 
described by the vector ${\bf \Phi}_x=(\Phi_x,\Phi_x')$.
In the symmetric case, $L_L=L_R=L'/2$, $M_L=M_R=M/2$, we obtain
\begin{eqnarray}
  {\bf M}_0 &=& \frac{1}{LL'-M^2}\left(\!\begin{array}{r r r r r}
      L' & L' & L' & -M &  M\\
      L' & L' & L' & -M &  M\\
      L' & L' & L' & -M &  M\\
      -M & -M & -M &  L & -L\\
       M &  M &  M & -L &  L
\end{array}\!\right)\!,\quad\quad\label{delft-M0}\\
  {\bf N} &=& \frac{1}{LL'-M^2}\left(\begin{array}{r r}
      -L' &  M\\
      -L' &  M\\
      -L' &  M\\
       M  & -L\\
      -M  &  L
\end{array}\right)\label{delft-N}
\end{eqnarray}
for the Hamiltonian and
\begin{equation}
  K(\omega) = \frac{i\omega}{Z(\omega)+ i\omega L'/4},
  \label{delft-K}
\end{equation}
$\mu=1/2$, ${\bf m} = (0,0,0,1,1)/\sqrt{2}$,
and ${\bf S} = -\sqrt{\mu}{\bf m}$.

Instead of quantizing the classical Hamiltonian Eq.~(\ref{Hamiltonian}), 
with Eqs.~(\ref{delft-M0}) and (\ref{delft-N}) we will linearize the dc SQUID in
order to separate the degrees of freedom that become very massive under the influence
of the external impedance from the other, light degrees of freedom.
Subsequently, we will only quantize the light degrees of freedom, viewing the
massive degrees of freedom as part of the environment.

\subsection{Linearization of the dc-SQUID}

We start by linearizing the uncoupled ($M=0$) SQUID.
The equations of motion for the SQUID are
\begin{eqnarray}
\frac{C'}{2}\left(\ddot\varphi_L-\ddot\varphi_R\right) &=& -\frac{1}{2L_J'}\left(\sin\varphi_L-\sin\varphi_R\right)\nonumber\\
&&-\frac{1}{L'}\left(\varphi_L-\varphi_R - 2\pi\frac{\Phi'_x}{\Phi_0}\right),\label{SQUID-diff}\\
\frac{C'}{2}\left(\ddot\varphi_L+\ddot\varphi_R\right) &=& -\frac{1}{2L_J'}\left(\sin\varphi_L+\sin\varphi_R\right) + \frac{2\pi}{\Phi_0}I_B\nonumber\\
&&- \mu K*\left(\varphi_L+\varphi_R\right)(t).\label{SQUID-sum}
\end{eqnarray}
Now we make the expansion
\begin{equation}
  \varphi_{L,R}(t) = \bar\varphi_{L,R} + \delta\varphi_{L,R}(t),
\end{equation}
where $\bar\varphi_{L,R}$ denotes the steady-state solution 
of the classical equations of motion Eqs.~(\ref{SQUID-diff}) and (\ref{SQUID-sum}).
We first find this steady-state solution in the absence of a bias current,
$I_B=0$, using $L'\ll L_J'$ and assume $\Phi_x'\neq \Phi_0/2$,
with the result
\begin{equation}
\bar\varphi_L^{(0)} = -\bar\varphi_R^{(0)} = \pi\frac{\Phi_x'}{\Phi_0}.
\end{equation}
Next, we allow a finite but small bias current $I_B \ll I_c'=\Phi_0/2\pi L_J'$,
and with $\bar\varphi_{L,R} = \bar\varphi_{L,R}^{(0)} + \delta\bar\varphi_{L,R}$
we find
\begin{equation}
\delta\bar\varphi_L = \delta\bar\varphi_R 
         = \pi\frac{L_J' I_B}{\Phi_0 \cos(\pi \Phi_x'/\Phi_0)}.
\label{IBcorrection}
\end{equation}
Starting from the steady-state solution, we can now derive the linearized
SQUID dynamics $\delta\varphi(t)$.  We assume
that the external impedance $Z(\omega)$ contains a sizable shunt capacitance 
$C_{\rm sh}\gg C'$ and that $\omega\ll1/\sqrt{L'C'}$ ($\approx 1500\,{\rm GHz}$
for typical values $C'=1\,{\rm fF}$, $L'=10\,{\rm pH}$).
Under these assumptions, the effect of the external 
impedance $Z\approx 1/i\omega C_{\rm sh}$ is to make the coordinate
$\varphi_L+\varphi_R$ very ``massive'', i.e.,
\begin{equation}
  K(\omega) \approx  \frac{i\omega C_{\rm sh}}{4} ,\quad\quad
  \int_0^t K(t-t') \varphi_+(t') \approx   \frac{C_{\rm sh}}{4}\ddot\varphi_+(t),
\end{equation}
the ``mass'' being $C'+C_{\rm sh}/4 \approx C_{\rm sh}/4$.
In order to eliminate $\varphi_L+\varphi_R$ from the classical equations
of motion, we introduce $\varphi_\pm=\varphi_L\pm\varphi_R$ 
and expand Eqs.~(\ref{SQUID-diff}) and (\ref{SQUID-sum}) about the
steady-state solution, $\varphi_\pm=\bar\varphi_\pm +\delta\varphi_\pm$,
\begin{equation}
\left[
\omega^2\frac{C'}{2}\!+\!
  \left(\!\!\begin{array}{c c}
      \frac{c_+}{2 L_J'}\!+\!\frac{1}{L'} & \frac{c_-}{2 L_J'}\\
      \frac{c_-}{2 L_J'} & \frac{c_+}{2 L_J'}\!+\! \mu K(\omega)
    \end{array}\!\!\right)
\right]
\!\left(\!\begin{array}{c}
      \delta\varphi_-(\omega)\\
      \delta\varphi_+(\omega)
    \end{array}\!\right)
\!=\!      \left(\!\begin{array}{c}
    0\\
    0
    \end{array}\!\right)\!,\nonumber\\
\label{SQUID-lin}
\end{equation}
where we have used the steady-state solution to define
\begin{eqnarray}
  c_+ &\equiv& \cos\bar\varphi_L + \cos\bar\varphi_R =  2\cos\pi\frac{\Phi_x'}{\Phi_0},\\
  c_- &\equiv& \cos\bar\varphi_L - \cos\bar\varphi_R = -2\pi\frac{L_J' I_B}{\Phi_0}\tan\pi\frac{\Phi_x'}{\Phi_0}.
\end{eqnarray}
Neglecting $C'\ll C_{\rm sh}$ in the equation of motion for $\delta\varphi_+$,
we can solve for $\delta\varphi_+$ (neglecting higher powers of $I_B$),
\begin{equation}
  \delta\varphi_+(\omega) = - c_- \left(c_+ + \frac{i \omega L_J'/2}{Z(\omega)+ i\omega L'/4}\right)^{-1} \delta\varphi_-(\omega).
\end{equation}
Substituting this back into Eq.~(\ref{SQUID-lin}), we obtain the following damping term in the equations of motion for $\delta\varphi_-$
\begin{equation}
\frac{c_-}{2 L_J'} \delta\varphi_+ (\omega) 
= -\frac{i\omega}{\tilde{Z}(\omega)}\delta\varphi_-(\omega),\label{SQUID-diss}
\end{equation}
with the effective SQUID inductance $\tilde{L}_J = L_J'/4\cos(\pi\Phi_x/\Phi_0) = L_J'/2c_+$
and the effective external impedance
\begin{equation}
  \tilde{Z}(\omega) = -\frac{\omega^2 L_J'^2}{Z_t(\omega)}\left(\frac{I_B}{I_c'}\tan\pi\frac{\Phi_x'}{\Phi_0}\right)^{-2},\label{Ztilde}
\end{equation}
where $I_c'$ is the critical current of the SQUID junctions and the total impedance 
(heavy SQUID degree of freedom in parallel with external impedance $Z$) is defined through
\begin{equation}
  Z_t(\omega) = \left(\frac{1}{i \omega \tilde{L}_J} + \frac{1}{Z(\omega)+ i\omega L'/4 }\right)^{-1}.\label{Zt}
\end{equation}
The effective external impedance $\tilde{Z}$ is much larger than $\omega L_J'$ for $I_B\ll I_c'$ or for $\sin\pi\Phi_x'/\Phi_0\approx 0$.
Thus, unlike $\delta\varphi_+$, the remaining degrees of freedom  (including $\delta\varphi_-$) 
are weakly affected by the effective external impedance and will be described as quantum 
mechanical degrees of freedom.

\subsection{Description of the light degrees of freedom}

\begin{figure}
\centerline{\includegraphics[width=7cm]{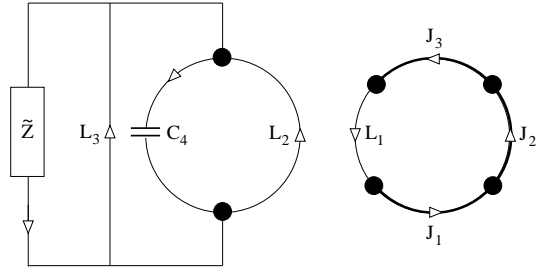}}
\caption{\label{delft-graph-red}
Circuit representation of the Delft qubit with a linearized SQUID.
The analysis is simplified due to the absence of tree inductors (K).}
\end{figure}

After having eliminated one degree of freedom from the SQUID, the remaining four degrees of
freedom $\bphi = (\varphi_1,\varphi_2,\varphi_3,\delta\varphi_-)$ will are now
described by the Hamiltonian Eq.~(\ref{Hamiltonian}) with the capacitances
${\bf C} = {\rm diag}(C,C,C,C'/2)$, the Josephson effective inductances
${\bf L}_{J}^{-1} = {\rm diag}(L_J^{-1},L_J^{-1},\beta L_J^{-1},0)$, and
\begin{eqnarray}
{\bf M}_0 &=& \frac{1}{L L'-M^2}\left(\begin{array}{r r r r}
    L' & L' & L' & -M \\
    L' & L' & L' & -M \\
    L' & L' & L' & -M \\
    -M & -M & -M & L    
\end{array}\right) \nonumber\\
&& +\frac{1}{\tilde{L}_J}\left(\begin{array}{r r r r}
    0 & 0 & 0 & 0\\
    0 & 0 & 0 & 0\\
    0 & 0 & 0 & 0\\
    0 & 0 & 0 & 1
\end{array}\right),\label{delft-red-M0}\\ 
  {\bf N} &=& \frac{1}{L L'-M^2}\left(\begin{array}{r r r r r}
      -L' &  M \\
      -L' &  M \\
      -L' &  M \\
       M &  -L
\end{array}\right).\label{delft-red-N}
\end{eqnarray}
Since the part of the circuit that was coupled to the bias current 
is described by $\tilde Z$, there is no coupling to a bias current left,
${\bf S}=0$.
By inspecting Eq.~(\ref{SQUID-diss}), we find
${\bf m}=(0,0,0,1)$, $\mu=1$, and
\begin{equation}
K(\omega) = \frac{i\omega}{\tilde{Z}(\omega)}.\label{delft-K-red}
\end{equation}
The results Eqs.~(\ref{delft-red-M0})--(\ref{delft-K-red}) for the reduced
system can also be obtained from the circuit drawn in Fig.~\ref{delft-graph-red}
with $C_4=C'/2$ and the inductance matrix
\begin{equation}
{\bf L} = \left(\begin{array}{c c c}
L & M  & 0\\
M & L' & 0\\
0 & 0  & \tilde L_J
\end{array}\right).
\end{equation}
Using Eqs.~(\ref{T1-dw}), (\ref{Tphi-dw}), (\ref{delft-K-red}), and (\ref{Ztilde}),
we obtain
\begin{eqnarray}
  \frac{1}{T_1} &=& \left(\frac{\Delta}{\omega_{01}}\right)^2\left(\frac{\Phi_0}{2\pi}\right)^2 |{\bf m}\cdot\Delta\bphi|^2 \frac{1}{\omega_{01}} {\rm Re}Z_t(\omega_{01})  \times\nonumber\\
&&\times\left(\frac{2\pi I_B}{\Phi_0}\tan\pi\frac{\Phi_x'}{\Phi_0}\right)^2   \coth\frac{\omega_{01}}{2k_B T},\label{T1-t}\\
  \frac{1}{T_\phi} &=& \left(\frac{\epsilon}{\omega_{01}}\right)^2\left(\frac{\Phi_0}{2\pi}\right)^2 |{\bf m}\cdot\Delta\bphi|^2 {\rm Re}Z_t(0) \times\nonumber\\
&&\times\left(\frac{2\pi I_B}{\Phi_0}\tan\pi\frac{\Phi_x'}{\Phi_0}\right)^2  2k_B T .\label{Tphi-t}
\end{eqnarray}

We make the approximation that the SQUID is completely classical;  solving the classical
equation of motion Eq.~(\ref{eq-motion-1}) for $\delta\varphi_-$ using
Eq.~(\ref{delft-red-M0}), we obtain the 
stationary classical solution for $\delta\varphi_-$ (with $L',L\ll L_J',\tilde L_J$),
\begin{equation}
  \delta\varphi_- = -\frac{2\pi}{\Phi_0}\left(M I +\Phi_x'\right),
\end{equation}
where we have used that $\sum_i\varphi_i = -2\pi\Phi/\Phi_0$, where 
$\Phi$ is the flux threading the qubit loop, and $\Phi-\Phi_x = L I$,
where $I$ is the current circulating in the qubit loop.
The difference between the two minima $\bphi$ 
(localized states $|0\rangle$ and $|1\rangle$) is then ($\Phi_x$ is constant)
\begin{equation}
  {\bf m}\cdot\Delta\bphi  = \Delta\delta\varphi_- \approx   -\frac{2\pi}{\Phi_0}M(I_L-I_R) ,
\end{equation}
and since $I_L=-I_R\equiv I$,
\begin{equation}
  {\bf m}\cdot\Delta\bphi \approx -2\frac{2\pi}{\Phi_0}MI .
\end{equation}
Substituting the above result into Eqs.~(\ref{T1-t}) and (\ref{Tphi-t}), we obtain
\begin{eqnarray}
  \frac{1}{T_1} &=& 4\left(\frac{\Delta}{\omega_{01}}\right)^2 I_B^2 \frac{1}{\omega_{01}} \left(\frac{2\pi M I}{\Phi_0}\tan\pi\frac{\Phi_x'}{\Phi_0}\right)^2\times\nonumber\\ &&\times 
{\rm Re}Z_t(\omega_{01}) \coth\frac{\omega_{01}}{2k_B T}\label{T1-TL}\\
  \frac{1}{T_\phi} &=& 4\left(\frac{\epsilon}{\omega_{01}}\right)^2 I_B^2 \frac{1}{\omega_{01}} \left(\frac{2\pi M I}{\Phi_0}\tan\pi\frac{\Phi_x'}{\Phi_0}\right)^2 \times\nonumber\\ &&\times
{\rm Re}Z_t(0) 2k_B T ,\label{T2-}
\end{eqnarray}
which agrees with earlier results \cite{WWHM}.
We also obtain an estimate for the leakage rate
from Eq.~(\ref{leakage-TL}),
\begin{equation}
  \label{delft-leakage}
  \frac{1}{T_L} = \left(\frac{\delta}{\eta}\right)^2 \!\!\!I_B^2 \frac{1}{\eta} \!\!\left(\frac{2\pi M I}{\Phi_0}\tan\pi\frac{\Phi_x'}{\Phi_0}\right)^2 \!\!\!{\rm Re}Z_t(\eta) \coth\frac{\eta}{2k_B T}.
\end{equation}

\subsection{Asymmetric SQUID}
\label{delft-asym}

Up to now we have assumed that the SQUID ring in the Delft qubit
is symmetric in two senses;  namely that both the self-inductances
of the left and right halves of the ring are identical and the
critical currents of the Josephson junctions in the left and right
halves of the SQUID ring are identical.  Both symmetries are
certainly broken to some degree in real physical systems.
Below, we study both cases, i.e.\ the case where the self-inductances
of the left and right halves of the ring are different (geometrical
asymmetry) and the case where the two critical currents
are different (Junction asymmetry).

\subsubsection{Geometric asymmetry}

We analyze the Delft qubit again with the inductance matrix
Eq.~(\ref{inductance-delft}) and the asymmetric inductances
\begin{eqnarray}
  M_L &=& \left(1+\frac{\alpha}{2}\right)\frac{M}{2},\quad
  M_R  =  \left(1-\frac{\alpha}{2}\right)\frac{M}{2},\\
  L_L &=& \left(1+\frac{\alpha}{2}\right)\frac{L'}{2},\quad
  L_R  =  \left(1-\frac{\alpha}{2}\right)\frac{L'}{2},
\end{eqnarray}
where $M_L+M_R=M$ and $L_L+L_R=L'$.
By linearizing the SQUID, we obtain the result
\begin{equation}
\tilde{Z}(\omega) = -\frac{\omega^2 L_J'^2}{Z_t(\omega)}\left[\left(\frac{I_B}{I_c'}\tan\pi\frac{\Phi_x'}{\Phi_0}\right)^2 + \alpha  \frac{I_B}{I_c'}\sin\pi\frac{\Phi_x'}{\Phi_0}\right]^{-1}.
\end{equation}
This result implies that if $\alpha \gg I_B/I_c'$, the decoherence rates
scale as $\alpha I_B/I_c'$ instead of $(I_B/I_c')^2$ for $\alpha \ll I_B/I_c'$.
Therefore, for very asymmetric loops, $\alpha \gg I_B/I_c'$, moderate
bias currents can already cause large decoherence effects.

\subsubsection{Junction asymmetry}

For asymmetric critical currents, or, equivalently,
asymmetric effective Josephson inductances,
\begin{equation}
  L_{J;L,R}' = L_J' (1\pm \gamma/2),
\end{equation}
we repeat the linearization of the SQUID keeping contributions of lowest order in $\gamma$.
Setting $I_B=0$, we obtain
\begin{equation}
  \frac{1}{L_{J;L,R}}\approx\frac{1}{L_J}\left(1\mp\frac{\gamma}{2}\right)+O(\gamma^2).
\end{equation}
The steady state of the SQUID is then determined by the
following equations,
\begin{eqnarray}
  \bar\varphi_L - \bar\varphi_R    =  -2\pi\frac{\Phi_x'}{\Phi_0}, &&\\
  \left(1-\frac{\gamma}{2}\right)\sin\bar\varphi_L 
  +\left(1+\frac{\gamma}{2}\right)\sin\bar\varphi_R &=& 0.
\end{eqnarray}
We make the ansatz
\begin{equation}
  \bar\varphi _{L,R} = \mp \pi\frac{\Phi_x'}{\Phi_0} + \gamma q,
\end{equation}
and obtain the result $q=-\tan(\pi\Phi_x'/\Phi_0)/2$, and finally,
\begin{equation}
  \bar\varphi _{L,R} = \mp \pi\frac{\Phi_x'}{\Phi_0} - \frac{\gamma}{2}\tan\pi\frac{\Phi_x'}{\Phi_0}.
\end{equation}
Comparing this to Eq.~(\ref{IBcorrection}), we see that in order to obtain
the Redfield tensor and the decoherence rates at zero bias current in
the presence of a junction asymmetry $\gamma$, we simply have to make the substitution
\begin{equation}
  \frac{I_B}{I_c'}\longrightarrow \gamma \sin\pi\frac{\Phi_x'}{\Phi_0}.
\end{equation}
Typical values for the junction asymmetry due to processing inaccuracies
are fairly large, $\gamma \approx 10\%$.
The effect of a junction asymmetry is more severe than the effect of a geometrical
asymmetry because for asymmetric junctions, decoherence occurs even for zero
bias current $I_B=0$.

\acknowledgments
We would like to thank Alexandre Blais for valuable discussions.
This work was supported in part by the National Security Agency, the Advanced Research
and Development Activity through Army Research Office contract number DAAD19-01-C-0056,
and by the DARPA QuIST program MDA972-01-C-0052.

\appendix

\section{Symmetry of ${\bf M}_0$}
\label{M0-sym}

In this Appendix, we prove that the matrix ${\bf M}_0$ defined in Eq.~(\ref{M0})
is always symmetric, ${\bf M}_0^T = {\bf M}_0$.  This property is required in order
to find a potential $U(\bphi)$ generating the force term $-{\bf M}_0\bphi$ in the
equation of motion.
We write ${\bf M}_0={\bf F}_{CL} {\bf V} {\bf W}^{-1} {\bf F}_{CL}^T$ with
\begin{eqnarray}
{\bf V} &=& \tilde{\bf L}_L^{-1}\bar{\bf  L} 
= \openone _L + {\bf X}\tilde{\bf L}_K \bar{\bf F}_{KL},\\
{\bf W} &=& {\bf L}_{LL}= \bar{\bf L}+{\bf F}_{KL}^T \tilde {\bf L}_K \bar{\bf F}_{KL},
\end{eqnarray}
with the off-diagonal block of ${\bf L}_{\rm t}$ from Eq.~(\ref{inductance-2}),
\begin{eqnarray}
  {\bf X}   &=& -{\bf L}^{-1}{\bf L}_{LK}\bar{\bf L}_K^{-1},\label{X}\\
  {\bf X}^T &=& -{\bf L}_K^{-1}{\bf L}_{LK}^T\bar{\bf L}^{-1},\label{XT}
\end{eqnarray}
and show that ${\bf V} {\bf W}^{-1}$ is symmetric, thus proving that ${\bf M}_0$ 
is symmetric.  Note that in Eq.~(\ref{XT}) we have used that ${\bf L}_{\rm t}$
is symmetric.  As a first step of our proof, we note that the symmetry of
${\bf V} {\bf W}^{-1}$,
\begin{equation}
  ({\bf V} {\bf W}^{-1})^T   \equiv  ({\bf W}^T)^{-1} {\bf V}^T = {\bf V} {\bf W}^{-1},
\end{equation}
is equivalent to the relation
\begin{equation}
  {\bf V}^T {\bf W}= {\bf W}^T {\bf V}.
\end{equation}
As a second step, we use Eq.~(\ref{FKLtilde}) to show
\begin{equation}
  {\bf V}^T {\bf W} -  {\bf W}^T {\bf V} 
  = \bar{\bf F}_{KL}^T\tilde{\bf L}_{K}^T\left( {\bf Y} - {\bf Y}^T \right)\tilde{\bf L}_{K}\bar{\bf F}_{KL},
\end{equation}
where ${\bf Y}=(\tilde{\bf L}_K^T)^{-1}-{\bf F}_{KL}{\bf X}$.
The third and last step of the proof is to show that ${\bf Y}$ is 
symmetric, i.e.\ ${\bf Y}={\bf Y}^T$. 
For this, we rewrite Eqs.~(\ref{LKtilde}) and (\ref{FKLtilde}) as 
\begin{eqnarray}
  \tilde{\bf L}_K &=& (\openone _K -{\bf L}_K \bar{\bf F}_{KL}{\bf X})^{-1}{\bf L}_K,\\
  \bar{\bf F}_{KL} &=& {\bf F}_{KL}-{\bf X}^T \bar{\bf L}.
\end{eqnarray}
Using these relations, we can show that 
${\bf Y} = {\bf L}_K^{-1} + {\bf X}^T \bar{\bf L}{\bf X}$
which is manifestly symmetric.  This concludes the proof that ${\bf M}_0$
is symmetric.

\section{Symmetry of ${\bf M}_d$}
\label{Md-sym}

From Eqs.~(\ref{eq-motion-0}), (\ref{sol-I}), and (\ref{sol-Z}), we obtain
hEq.~(\ref{eq-motion-1}) with
\begin{widetext}
\begin{eqnarray}
  {\bf M}_d(\omega)   &=&  {\bf F}_{CL}\left(\tilde {\bf L}_L^{-1} \bar {\bf L}{\bf L}_{LL}^{-1} {\bf L}_{LZ}- {\bf L}^{-1}{\bf L}_{LK}\bar{\bf L}_K^{-1}\tilde{\bf L}_K{\bf F}_{KZ}\right)\bar{\bf L}_{Z}^{-1}(\omega){\bf L}_{ZL}{\bf L}_{LL}^{-1}{\bf F}_{CL}^T \label{Md-full}\\
&& + {\bf F}_{CZ}\bar{\bf L}_Z^{-1}(\omega)\left({\bf F}_{CZ}^T - {\bf L}_{ZL}{\bf L}_{LL}^{-1}(\omega){\bf F}_{CL}^T\right) %\nonumber\\ &&
 + {\bf F}_{CL}\left({\bf L}^{-1}{\bf L}_{LK}\bar{\bf L}_K^{-1}\tilde{\bf L}_K{\bf F}_{KZ}
- \tilde{\bf L}_L^{-1}\bar {\bf L}{\bf L}_{LL}^{-1}{\bf L}_{LZ}\right)\bar{\bf L}_Z^{-1}(\omega){\bf F}_{CZ}^T,\nonumber
\end{eqnarray}
\end{widetext}
where we have used the identity 
$\bar{\bf L}_{L}^{-1}(\omega){\bf L}_{LZ}{\bf L}_{ZZ}^{-1}(\omega) 
= {\bf L}_{LL}^{-1}{\bf L}_{LZ}\bar{\bf L}_Z^{-1}(\omega)$.  
This expression is a quadratic form in ${\bf F}_{CL}$ and ${\bf F}_{CZ}$,
\begin{equation}
   {\bf M}_d = \left(\begin{array}{r r} \!\! {\bf F}_{CL} \! & \! {\bf F}_{CZ}\!\!\end{array}\right) 
                       \left(\begin{array}{r r} {\bf A}\bar{\bf L}_Z^{-1}{\bf B} & 
                                                {\bf A}\bar{\bf L}_Z^{-1}\\
                                                \bar{\bf L}_Z^{-1}{\bf B}        & 
                                                \bar{\bf L}_Z^{-1}\end{array}\right)
                       \left(\begin{array}{r}   {\bf F}_{CL}^T \\ {\bf F}_{CZ}^T\end{array}\right),  
\end{equation}
with the definitions
\begin{eqnarray}
{\bf A} &=& -\tilde{\bf L}_L^{-1}\bar{\bf L}{\bf L}_{LL}^{-1}{\bf L}_{LZ}
            +{\bf L}^{-1}{\bf L}_{LK}\bar{\bf L}_K^{-1}\tilde{\bf L}_K{\bf F}_{KZ},\\
{\bf B} &=& -{\bf L}_{ZL}{\bf L}_{LL}^{-1}
         =  -{\bf F}_{KZ}^T\tilde{\bf L}_K\bar{\bf F}_{KL}{\bf L}_{LL}^{-1}.
\end{eqnarray}
Next, we show that ${\bf M}_d$ must be symmetric, ${\bf M}_d^T={\bf M}_d$, and therefore
$\bar{\bf L}_Z^T=\bar{\bf L}_Z$ and ${\bf A}={\bf B}^T$.

The argument for the symmetry of ${\bf M}_d$ is as follows.
We consider a generalized model in which the external impedances $Z$ and
the linear inductances $L$ and $K$ are treated on an equal footing.
For this purpose, we allow mutual impedances (generalized mutual inductances)
between $Z$ and $K$ and include $Z$ into $L$ by allowing frequency dependent
linear inductances and writing $L_Z(\omega)=Z(\omega)/i\omega$.
This leaves us with the following types of circuit elements;
tree elements are either capacitors $C$ or linear impedances $K$ where
${\bf L}_K(\omega)={\bf Z}_K(\omega)/i\omega$, 
branch elements are Josephson junctions
(non-linear inductors) $J$, linear impedances $L$ where
${\bf L}(\omega)={\bf Z}_L(\omega)/i\omega$, and external bias currents $B$.
In addition to this, there can be frequency-dependent linear mutual impedances
${\bf Z}_{LK}(\omega)$, where ${\bf L}_{LK}(\omega)={\bf Z}_{LK}(\omega)/i\omega$,
between the $L$ and $K$ branches.
The equation of motion (\ref{eq-motion-1}) can now
be derived exactly as before, but in the frequency domain,
the result being Eq.~(\ref{eq-motion-1}) without the ${\bf M}_d(\omega)$
term, since there are no $Z$ branches.
These new equations include dissipation 
which is described by the (now frequency-dependent)
${\bf M}_0'(\omega)$, the prime distinguishing it from the ``ordinary''
${\bf M}_0$ (see above).  The matrix ${\bf M}_0'$ is formally identical to ${\bf M}_0$, 
up to frequency dependencies which are irrelevant for the symmetry
of the matrix.
We have shown in Appendix \ref{M0-sym} that ${\bf M}_0^T={\bf M}_0$;
this proof also goes through for ${\bf M}_0'$, thus
${\bf M}_0'^T={\bf M}_0'$.  
Since ${\bf M}_0'(\omega)={\bf M}_0+{\bf M}_d(\omega)$
and both ${\bf M}_0$ and ${\bf M}_0'$ are symmetric, 
we conclude that also ${\bf M}_d^T={\bf M}_d$.
Introducing $\bar{\bf m}={\bf F}_{CL}+{\bf F}_{CZ}{\bf A}
={\bf F}_{CL}+{\bf F}_{CZ}{\bf B}^T$, 
we can now write ${\bf M}_d$ in the form given in
Eqs.~(\ref{Md})~and~(\ref{mbar}).

\end{document}